\documentclass[]{aastex631}
\usepackage[normalem]{ulem}
\usepackage{textcomp, gensymb}
\useunder{\uline}{\ul}{}

\begin{document}

\title{Analysis of SEP events and their possible precursors based on the GSEP Catalog}

\correspondingauthor{Sumanth A. Rotti}
\email{srotti@gsu.edu}

\author[0000-0003-1080-3424]{Sumanth A. Rotti}
\affiliation{Georgia State University \\
Department of Physics and Astronomy \\
Atlanta, GA, USA}

\author{Petrus C. Martens}
\affiliation{Georgia State University \\
Department of Physics and Astronomy \\
Atlanta, GA, USA}


\begin{abstract}

Solar energetic particle (SEP) events are one of the most crucial aspects of space weather. Their prediction depends on various factors including the source solar eruptions such as flares and coronal mass ejections (CMEs). The Geostationary Solar Energetic Particle (GSEP) Events catalog was developed as an extensive data set towards this effort for solar cycles 22, 23 and 24. In the present work, we review and extend the GSEP data set by; (1) adding ``weak" SEP events that have proton enhancements from 0.5 to 10 pfu in the E$>$10 MeV channel, and (2) improving the associated solar source eruptions information. We analyze and discuss spatio-temporal properties such as flare magnitudes, locations, rise times, and speed and width of CMEs. We check for the correlation of these parameters with peak proton fluxes and event fluences. Our study also focuses on understanding feature importance towards the optimal performance of machine learning (ML) models for SEP event forecasting. We implement random forest (RF), extreme gradient boosting (XGBoost), logistic regression (LR) and support vector machines (SVM) classifiers in a binary classification schema. Based on the evaluation of our best models, we find both the flare and CME parameters are requisites to predict the occurrence of an SEP event. This work is a foundation for our further efforts on SEP event forecasting using robust ML methods.

\end{abstract}



\section{Introduction} \label{sec:intro}

High-energy particles are ejected out of the solar corona by shock waves \citep{gopalswamy2001predicting} and due to the release of significant amounts of magnetic energy in solar active regions during the reconnection process \citep{lin}. One aspect of such ejections is the solar energetic particles (SEPs) that predominantly constitute protons along with electrons and other ions. SEPs are accelerated from the Sun to the interplanetary medium by eruptive phenomena such as solar flares (SFs) and shocks followed by coronal mass ejections (CMEs) \citep{1986cane, 1992kahler,reames1999particle}. SFs can cause radio blackouts, and CMEs heading toward Earth with southward pointed magnetic field (in the opposite direction of Earth’s field at the Sun-facing side) can lead to geomagnetic storms \citep{pulkkinen2007space, aparna2020solar}. Although SEP events are rare in numbers compared to SFs and CMEs \citep{klein2001origin}, they have significant impacts in terms of space weather that includes different levels of technological disruptions \citep{smart1992} and biological perils on various economic scales \citep{schrijver2010heliophysics}. Typical health risks include severe radiation hazards to astronauts taking spacewalks and airline travels on polar routes \citep{beck2005tepc, schwadron2010earth, jiggens2019situ}. According to the Space Weather Prediction Center (SWPC), proton intensities $\ge$10 pfu (1 pfu = 1 particle per cm$^2$.s.sr) in the E$>$10 Mega electron-Volt (MeV) energy channel are termed as large SEP events with regards to causing significant space weather effects \citep{bain2021summary}.

With great advancements in space engineering and technology, space-based satellites have acquired near-continuous observations of SFs and CMEs in multiple wavelengths since 1997. For SEPs, \textit{in situ} measurements show characteristic behaviors with energetic proton flux profiles rising to peak values in a few minutes to several hours \citep{kallenrode2003current, klein2005onset, kahler2005characteristic, cane2006introduction}. We will show further in our discussions how these flux profiles depend on the energy of protons, and the Sun-Earth connection. All these solar data have provided insights to infer a wide range of space weather consequences from solar eruptions \citep{gopalswamy2003effect}. One important aspect of analyzing solar data is to advance operational capabilities by mitigating space weather effects on our technological systems \citep{jackman1987solar}. This requires the development of a robust tool to predict eruptive event occurrences.

Statistical analyses of SEP event properties have been carried out for decades by several researchers \citep{1975van, laurenza2009technique, dierckxsens2015relationship, papaioannou2016solar}. A principal component analysis (PCA) considering a combination of six solar variables (namely, the CME width/size and velocity, SF longitude, duration and rise time, and logarithm of the SF magnitude) was presented in \cite{papaioannou2018nowcasting}. The association of SEP events to a parent flare or CME has given us insights into particle acceleration and propagation in the interplanetary medium. Since SEP events depend on source solar eruptions, gathering statistical evidence of their causation and acceleration is vital for the forecasting of space weather consequences. One operational system that uses parameters of both SFs and CMEs to predict the occurences of SEP events is the Forecasting Solar Particle Events and Flares (FORSPEF) tool (see \cite{anastasiadis2017predicting} for details.) In addition, great efforts are being made toward implementing the proven abilities of machine learning (ML) techniques to predict solar eruptive events \citep{enrico2019}. Efforts on SEP event forecasting based on ML have been of a crucial focus globally with upcoming human missions to Moon and Mars \citep{WHITMAN2022}. Best practices in ML comprise building robust models on high-quality datasets, so-called benchmarks. The advantage of a carefully curated data set is that, with the right computational tools, it provides clues to understand and explore eruptive activities, during various phases (rise, peak, and descend) of the solar cycle.

Existing SEP event forecasting tools exploit the characteristics of source solar events \citep{posner2007up}. Studies based on flare and CME parameters individually or a combination of both have been made in the last two decades \citep{2007kahler, 2011nunez, falconer2011tool, 2015nunez}. The earliest evidence of two distinct processes of particle acceleration leading to SEP events was obtained from radio observations \citep{1963Wild}. Over the last three decades, several researchers have put efforts into exploring SEP-associated slow drift rate type-II and fast drift rate type-III radio bursts statistically \citep{Nicholson1978ApJ, gopalswamy2002interacting, Cliver2004ApJ, balch, cliver_ling2009ApJ, laurenza2009technique, winter2015ApJ, alberti2017solar}.

In this work, we analyze and describe statistical correlations of SEP events with associated solar source eruptions. We explore the properties of SEP events from solar cycles 22, 23, and 24. We consider the times and peak intensities of flare fluxes, CME speed and angular width as they are considered to influence the SEP event populations. The longitudinal dependence of particle propagation from the solar surface is also studied. Although real-time radio data is unavailable, we briefly highlight its relevance as a vital element in forecasting efforts. Nonetheless, we defer our analysis of radio data to future work. We study feature importance of solar source parameters using two tree-based classifiers, namely,  random forests and extreme gradient boosting models. We also evaluate the performance of our best models in a supervised binary classification schema on an undersampled subset of our data. This ML experiment forms a vital step to our future efforts in building robust models for SEP event forecasting.

Our focus in the present study is to (1) provide an updated data set of SEP events; (2) explore the combined effects of including flare and CME parameters towards forecasting SEP events using a carefully curated benchmark data set. In this work, we consider the term `SEP events' analogous to solar protons events (SPEs). The data set used in this study is described in section \ref{sec:data}. We discuss the methodologies and techniques used for source-event association in section \ref{sec:analysis}. We analyze the properties of solar source eruptions associated to SEP events in section \ref{sec:prop}. We present our findings and results in section \ref{sec:results} discussing the SEP event properties, correlations of SEP and source parameters, the feature importance study based on tree-based ML models, and the binary classification techniques implemented in our work. Lastly, in section \ref{sec:conclusion}, we present the conclusions from our work.

\section{Data Sources} \label{sec:data}

The Space Environment Monitor (SEM) suite \citep{grubb} on board the Geostationary Operational Environmental Satellite (GOES) \citep{sauer1989sel, bornman} obtain \textit{in situ} measurements of SEPs and are made publicly available by the National Oceanic and Atmospheric Administration (NOAA) in their online database\footnote{\url{https://www.ngdc.noaa.gov/stp/satellite/goes/index.html}}. We follow the definition of NOAA for a significant SEP event as crossing the 10 pfu threshold in the P3 channel. We consider classifying the SEP events into two categories based on the NOAA-SWPC threshold. That is, based on the integral proton flux ($I_{P}$) recorded by P3, we define as “large” those events for which $I_{P}$$\geq$10 pfu and “weak” events those for which 0.5$<$$I_{P}$$<$10 pfu.

Our recently published geostationary solar energetic particle (GSEP) events catalog\footnote{The GSEP data set available on Harvard Dataverse: \dataset[10.7910/DVN/DZYLHK]{https://doi.org/10.7910/DVN/DZYLHK}} has been the fundamental source of the present work. The GSEP data set stands as a large database of SEP events for solar cycles 22, 23 and 24 to the solar physics community. It was developed with multiple aspects: (1) integration of SEP events from existing databases, (2) identification and association of source solar eruptions with each SEP event, (3) including metadata that consists of physical parameters relevant to source eruptions, and (4) providing cleaned time series slices for each SEP event. In this work, we present the updates and improvements that were necessary to be made in the GSEP catalog in all above aspects which we discuss further.

The primary part of the current work has been identifying and adding ``weak" events to the GSEP data set. The description of our extended data set development follows the methodology of \citet{rotti2022}. The integrated E$\geq$10 MeV proton fluxes corresponding to the P3 channel in the GOES-SEM have been used in this study. In the Appendix, table \ref{tab:goes} lists the GOES series from 1986 to 2017 considered in the development of the GSEP data set. Here, the choice of the GOES missions depends on continuous data availability with fewer gaps and errors. In most cases, we considered the GOES proton data that recorded the largest intensity. However, exceptions are for inaccurate/invalid data where we visually verified measurements from each SEM instrument making parallel observations. In addition to proton fluxes, we utilize the one-minute averaged GOES soft x-ray (1–8Å) fluxes measured by the X-Ray Sensor (XRS) onboard GOES. The archived data is available online from NOAA's website\footnote{\url{https://www.ncei.noaa.gov/data/goes-space-environment-monitor/access/avg/}}.

When there are uncertainties in solar source associations for SEP events, we examine observations ranging from extreme ultraviolet (EUV) to X-ray images and data from the Solar and Heliospheric Observatory (SOHO) mission's coronagraphs \citep{brueckner1995}. To examine and associate source solar eruptions, we utilize: (1) the GOES flare catalog\footnote{\url{https://www.ngdc.noaa.gov/stp/space-weather/solar-data/solar-features/solar-flares/x-rays/goes/xrs/}}; (2) the CDAW CME catalog\footnote{\url{https://cdaw.gsfc.nasa.gov/CME_list/}}, and (3) the CDAW type-II radio burst catalog\footnote{\url{https://cdaw.gsfc.nasa.gov/CME_list/radio/waves_type2.html}}. The respective CDAW catalogs have been available since 1997 only. Hence, no data on CMEs and radio bursts have been included for cycle 22 in our data set. Also, we have not surveyed radio events observed from the ground.

\begin{figure}[ht!]
\plotone{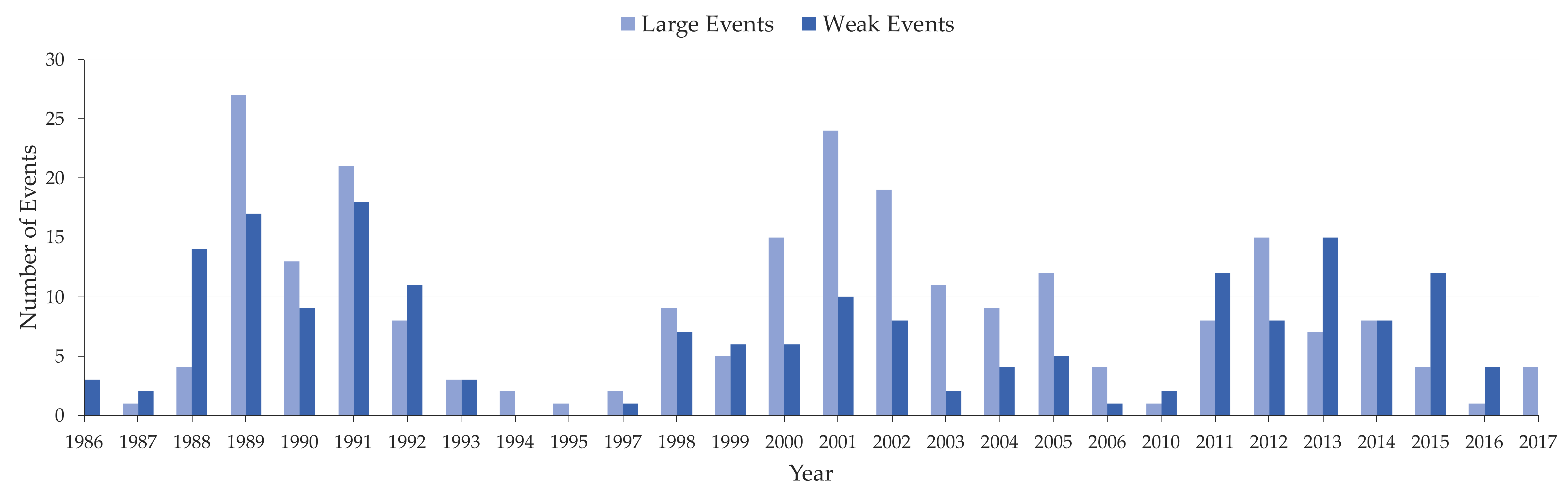}
\caption{This plot shows the strength of solar activity in terms of SEP events between 1986 to 2017. A total of 163, 160, and 110 events have been identified in cycles 22, 23, and 24, respectively. In the legend, ``large" events are defined as those crossing the SWPC threshold of 10 pfu in the E$\geq$10 MeV channel, and ``weak" events are $\geq$0.5 pfu but $<$10 pfu.
\label{fig:sep_sc}}
\end{figure}

\section{Data Analysis} \label{sec:analysis}
The severity of the proton events is measured using the NOAA Solar Radiation Storm Scale (S-scale)\footnote{\url{https://www.swpc.noaa.gov/noaa-scales-explanation}} that relates to biological impacts and effects on technological systems. The S-scale relies on the E$\geq$10 MeV integral peak proton flux that characterizes the intensity of an SEP event. The base threshold, designated as `S1', corresponds to a GOES 5-min averaged E$\geq$10 MeV integral proton flux exceeding 10 pfu for at least three consecutive readings \citep{bain2021summary}. Further scales from `S2' to `S5' logarithmically increase from one another, therefore defining different event sizes.

Our data set consists of 433 events, wherein 244 (189) are large (weak) in terms of SEP event intensity. Some events within data gaps of GOES observations have been verified by cross-referring SEP events from other catalogs (see table 1 in \cite{rotti2022}) and interpolating the missing data. In certain scenarios related to SEPs, a passing interplanetary shock causes energetic storm particle (ESP) acceleration \citep{CANE199535}. Nine events in our data set during the solar maximum are ambiguous in particle production, primarily due to the injection of fluxes from previous events. By visual inspection of the GOES proton data, a rise in the fluxes is noticed after the passage of a shock. The most reliable convention was suggested by the Space Radiation Analysis Group (SRAG) to consider such follow-up events as ESPs (private communications with Dr. Steve Johnson).

Figure \ref{fig:sep_sc} shows the distribution of all the SEP events in our data set from 1986 to 2017. It can be noticed that the amplitude of the cycles in terms of SEP events has decreased over the last three cycles, although, cycles 22 and 23 are almost equal in strength while SC24 is less active. The percentage ratio between large and weak events is as follows:

1.	SC22 – 52:48

2.	SC23 – 68:32

3.	SC24 – 44:56

Because of NOAA's definition, there are a few events in our data set that are very close to being considered large SEP events. We observe eight such events to have peak proton fluxes fluctuating between 10-12 pfu in the E$\geq$10 MeV channel. Similarly, there are twelve events with peaks oscillating between 12-13 pfu. The proton measurements in all the scenarios are above the 10 pfu threshold for three consecutive points. However, on estimating the fluence levels we notice they are below the threshold of a radiation hazard. Hence, we consider these 20 cases as ``weak" SEP events.

We have put effort into identifying for each SEP event in our data set with at least one eruption, either a flare or CME. As we will show later on in the manuscript, a probable precursor occurs within 12 hours of SEP onset for many events in our data set. The temporal profiles of SEP events positively correlate with longitude, with a few exceptions. Eastern hemisphere events typically have gradually rising proton profiles, whereas western events reach peak fluxes within a few hours of the parent flare eruption. We iteratively identified consistent source associations for exceptional SEP events. However, nine SEP events in our data set could not be associated with any solar source information. This probably constitutes a few far-side events that lack observational evidence to match with a source eruption.

\subsection{Solar source selection}
For most cases in our data set, the association of an SEP event with a source is straightforward. Many active regions giving rise to a flare or CME are distributed on the visible disc of the Sun while a few are on the backside. For those beyond limb and backside events, an estimate of the likely source active region has been made in reference to existing methods. A simple technique was to follow the heliographic longitudes of active regions on the visible disk that were previously sources of intense flares and SEPs.

\begin{figure*}[ht!]
\gridline{\fig{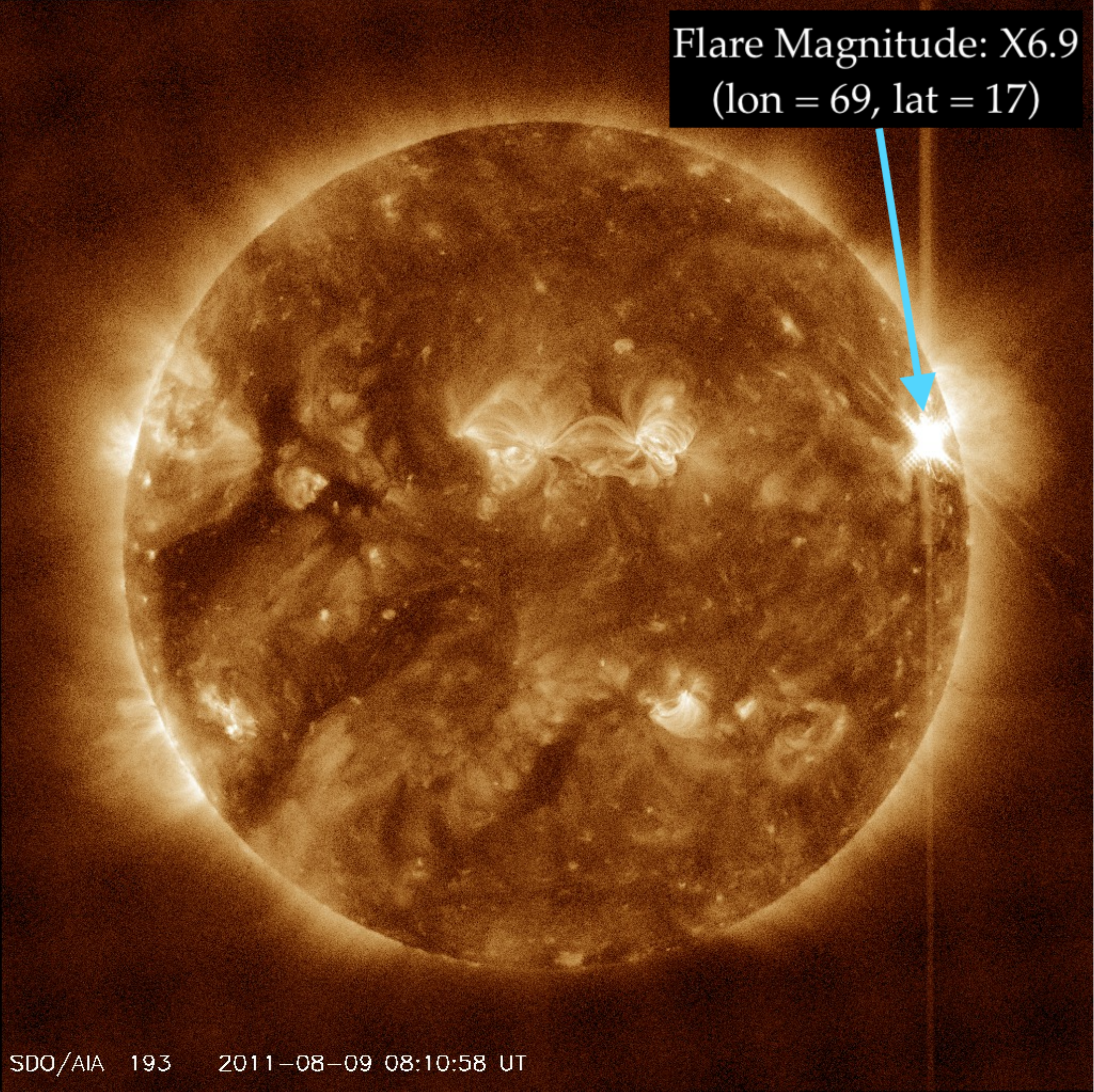}{0.45\textwidth}{(a)}
          \fig{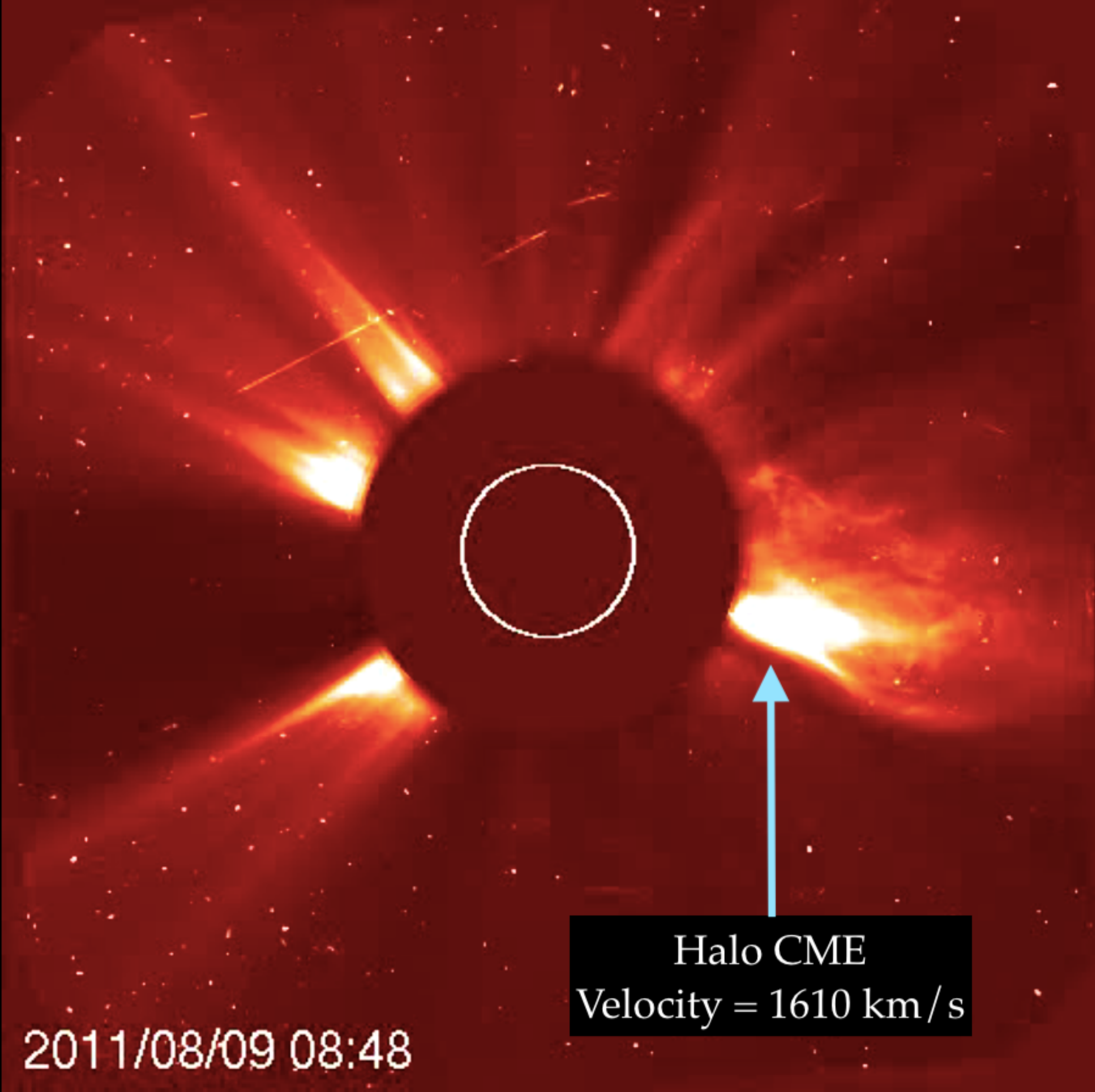}{0.45\textwidth}{(b)}
          }
\gridline{\fig{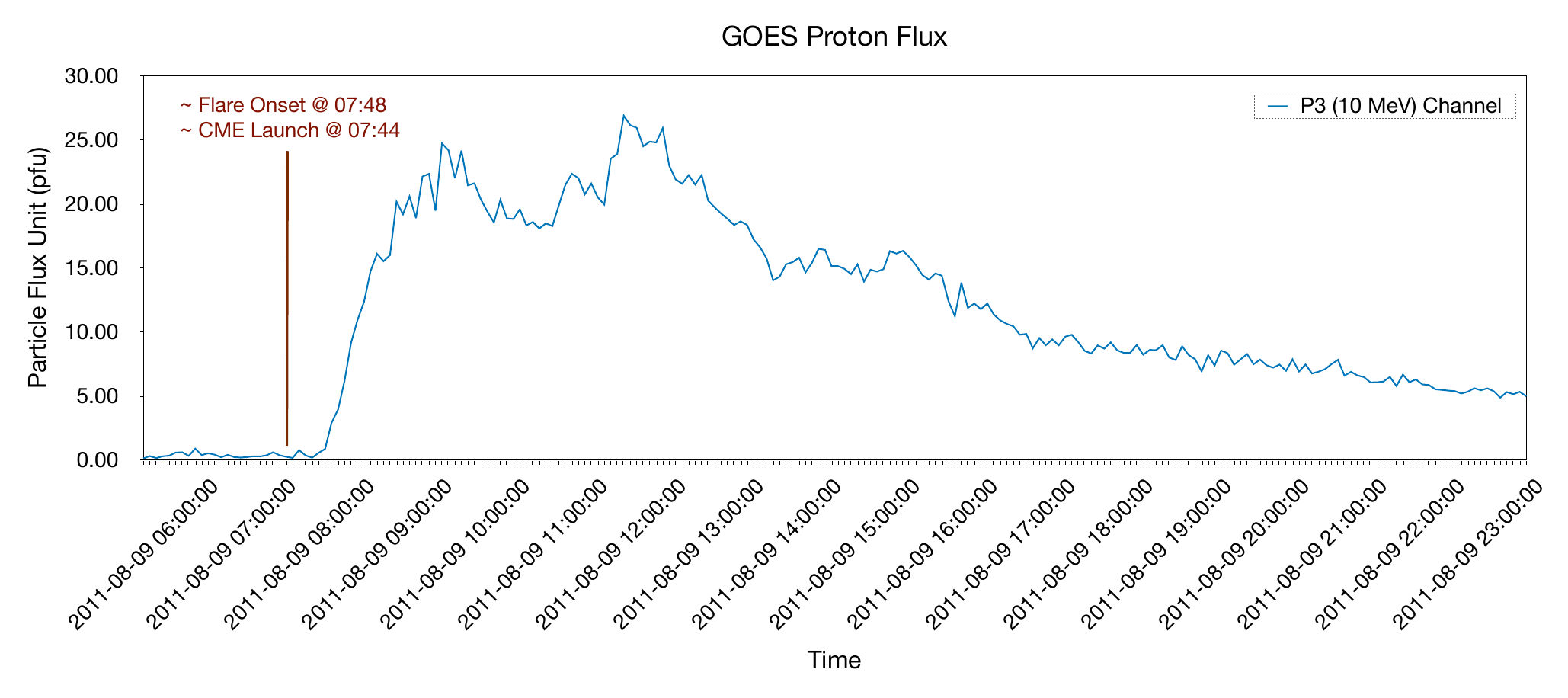}{0.9\textwidth}{(c)}
          }
\caption{Plots of parent eruptions followed by an SEP event on 2011-08-09 indexed as {\tt\string gsep$\_{292}$} in the GSEP data set. (a) A solar flare of magnitude X6.9 seen in in 193{\AA} erupted from active region 11263 with spatial extents: lon = 69, lat = 17. \textit{Image courtesy}: SDO/AIA. (b) A halo CME traveling at a mean speed of $\approx$ 1610 km.s$^{-1}$ was launched less than four minutes after the parent flare. \textit{Image courtesy}: SOHO/LASCO-C2. (c) Proton intensity profile plot of the SEP event with a peak flux of 26 pfu in the E$\geq$10 MeV integral channel measured by the GOES-SEM suite.
\label{fig:ex1}}
\end{figure*}

Figure \ref{fig:ex1} shows the parent eruptions and associated and SEP event on 2011-08-09 indexed as {\tt\string gsep$\_{292}$} in the GSEP data set. In \ref{fig:ex1}(a) is the image of the Sun showing an instance of the source flare eruption taken in 193{\AA} by the Atmospheric Imaging Assembly (AIA) onboard the Solar Dynamics Observatory (SDO) \citep{aia}. The flare originated from active region 11263 (lon = 69, lat = 17) and had a magnitude of X6.9 as measured by the GOES/XRS instrument. Following the flare was a halo fast-CME propagating with a velocity of $\approx$1610 km.s$^{-1}$ shown in \ref{fig:ex1}(b) as observed by LASCO/C2 instrument several minutes after the launch time. In the frame shown, we can see the evolving CME cloud (bright region) out of the solar corona and the shock-front is visible at the side of the CME. There was also a DH type-II radio burst observed $\sim$35\ mins after the CME eruption. In \ref{fig:ex1}(c) is the time series plot of the subsequent large SEP event with a peak flux of 26 pfu in the E$\geq$10 MeV channel measured by the GOES-SEM instrument. The vertical line overlayed in the plot indicates an approximate start time of the parent eruptions. The onset of this SEP event occurs at 08:45 with a rise time of three hours and 25 minutes. The event lasted for more than eight hours until it subsided below the threshold of 10 pfu in the P3 channel.

\begin{figure*}[ht!]
\gridline{\fig{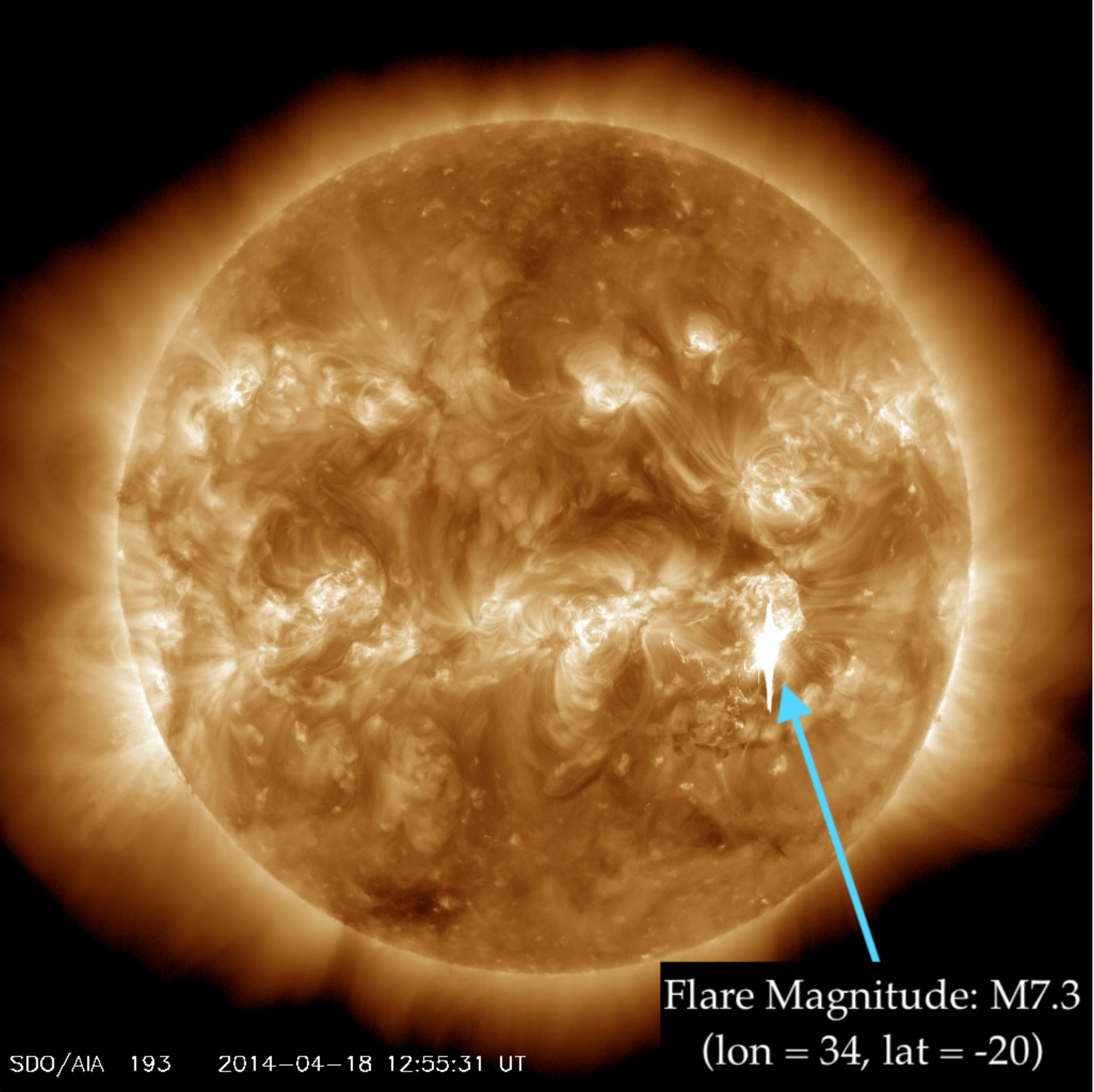}{0.45\textwidth}{(a)}
          \fig{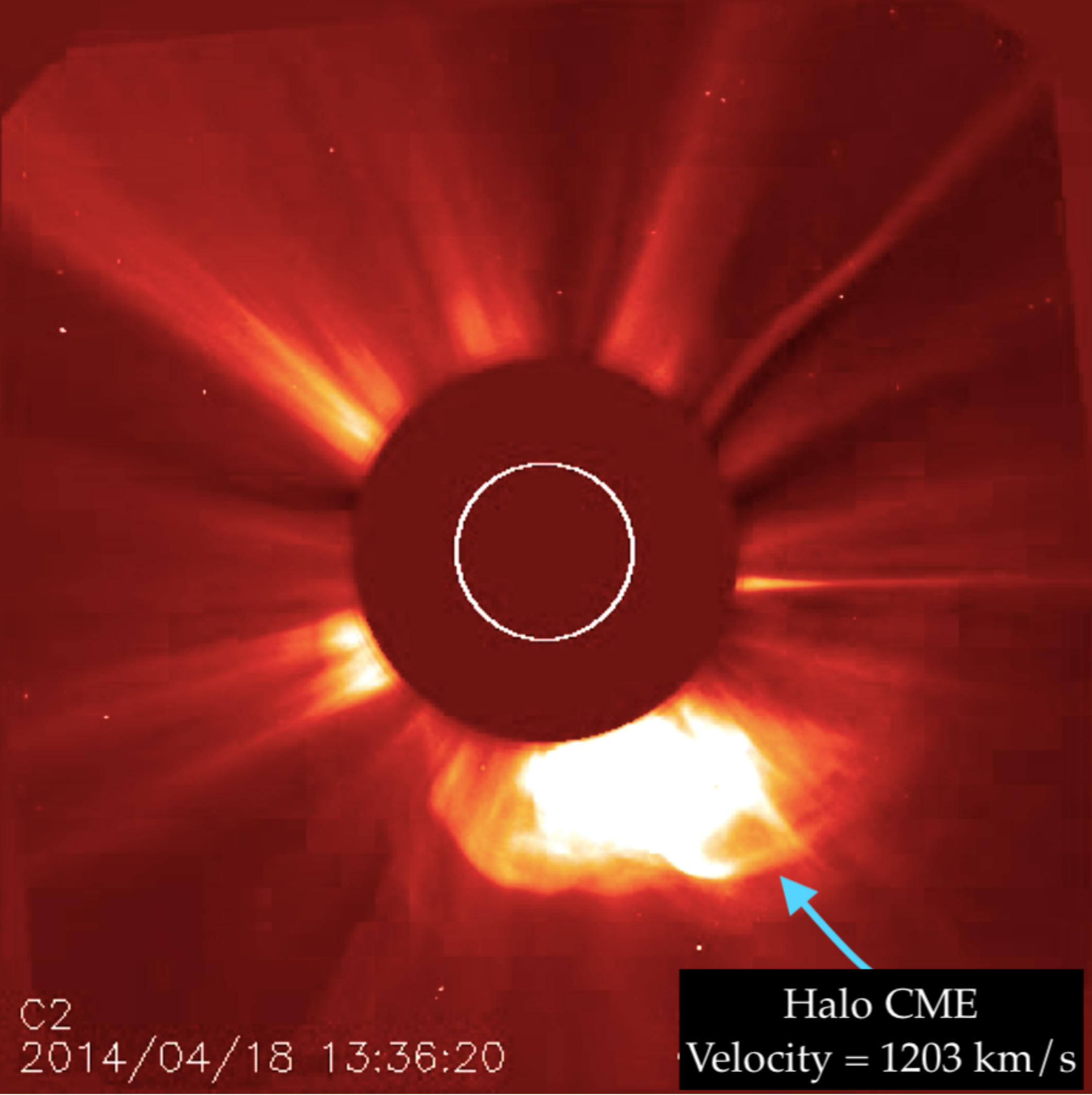}{0.45\textwidth}{(b)}
          }
\gridline{\fig{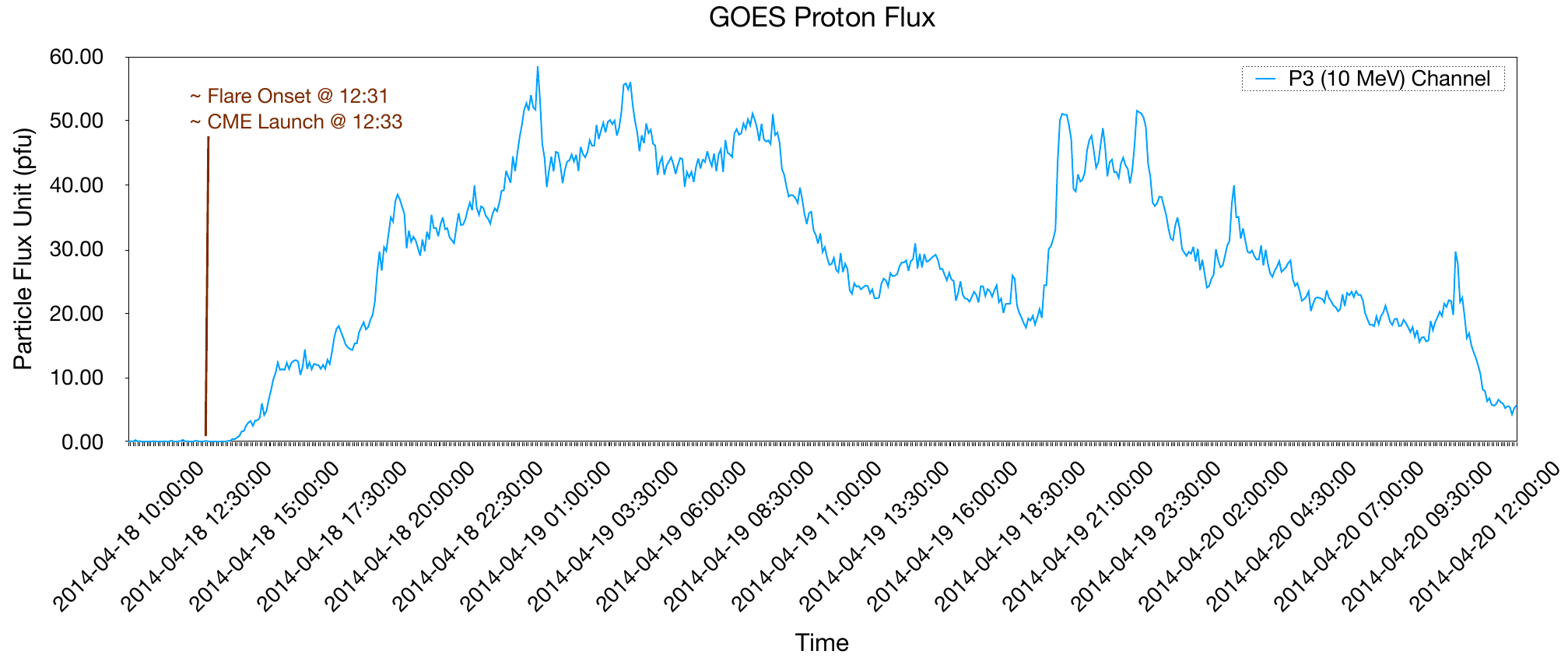}{0.9\textwidth}{(c)}
          }
\caption{Plots of parent eruptions followed by an SEP event on 2014-04-18 indexed as {\tt\string gsep$\_{330}$} in the GSEP data set. (a) A solar flare of magnitude M7.3 erupted from active region 12033 with spatial extents: lon = 34, lat = -20. \textit{Image courtesy}: SDO/AIA. (b) A halo CME traveling at a mean speed of $\approx$ 1203 km.s$^{-1}$ was launched less than three minutes after the parent flare. \textit{Image courtesy}: SOHO/LASCO. (c) Proton intensity profile plot of the SEP event with a peak flux of 58 pfu measured by the GOES-SEM instrument in the E$\geq$10 MeV integral channel.
\label{fig:ex2}}
\end{figure*}

Figure \ref{fig:ex2} shows the parent eruptions and associated SEP event on 2014-04-18 indexed as {\tt\string gsep$\_{330}$} in the GSEP data set. In \ref{fig:ex2}(a) is the image of the Sun showing an instance of the source flare eruption taken in 193{\AA} by the SDO/AIA instrument. The flare erupted less than three hours before the SEP event onset from active region 12033 (lon = 34, lat = -20) and had a magnitude of M7.3 as measured by the GOES/XRS instrument. Following the flare was a halo fast-CME propagating with a velocity of $\approx$ 1203 km.s$^{-1}$ shown in \ref{fig:ex2}(b) as observed by LASCO/C2 instrument several minutes after the launch time. In the frame shown, we can see the evolving CME cloud (bright region) out of the solar corona and the shock-front is visible around the CME. There was also a DH type-II radio burst associated with this CME. Figure \ref{fig:ex2}(c) shows the time series plot of the subsequent large SEP event with peak flux of 58 pfu in the E$\geq$10 MeV channel measured by the GOES-SEM instrument. The vertical line overlayed in the temporal plot approximately indicates the start time of the parent eruptions. The onset of this SEP event occurs at 15:25 (UT) with a rise time of nine hours and 40 minutes.

\subsection{Metadata}
The metadata in the updated GSEP catalog consists of corrected and carefully associated source eruptions. In the latest version, proton fluence estimates are provided for each SEP event. We also include a ``Flag" column to indicate ``1'' or ``0'' representing ``large" or ``weak" events, respectively. Complete source solar eruption information is available for as many as 144 SEP events in our data set. In table \ref{tab:lists}, we present the total number of possible solar sources for each SEP event. 

\begin{deluxetable*}{lcccccc}[ht!]
\tablenum{1}
\tablecaption{Total SEP events and their solar sources in the GSEP catalog. \label{tab:lists}}
\tablewidth{0pt}
\tablehead{
\textbf{}   &  \multicolumn{3}{c}{Large Events ($\geq$10pfu)} & \multicolumn{3}{c}{Weak Events ($<$10pfu)}    \\
\colhead{} & \colhead{} & \colhead{} &
\colhead{} & \colhead{} & \colhead{} \\
\cline{2-7}
                        & \textbf{SC22} & \textbf{SC23} & \textbf{SC24} & \textbf{SC22} & \textbf{SC23} & \textbf{SC24} \\
}
\startdata
SEP Events    &  86       & 110      & 48         & 77           & 50   & 62                   \\
Active Region Data       &  78       & 97       & 46          & 65             & 43 & 52                      \\
Flares      &  84    & 97            & 43               & 73             & 45 & 44                      \\
CMEs     &  -       & 97         & 48               & -           & 41 & 61                     \\
Type-II Radio Burst   & -      & 75    & 35               & -             & 21 & 16                      \\
\enddata
\tablecomments{The values in parenthesis in the header row denote the threshold of proton peak fluxes to classify SEP events into large and weak. There are no \textit{in situ} observations of CMEs and radio bursts for solar cycle 22. Hence, they are left blank in the respective rows. }
\end{deluxetable*}

A summary of our data set is as follows:

\begin{enumerate}
    \item We obtain 386 cases of flare information for associated SEP events.
    \item There are 16 flaring events without an associated CME and active region data.
    \item We notice 18 CMEs with no supportive observational relations to flares and active regions.
    \item Three CME events have no information about their widths and speeds.
    \item There are 147 type-II radio bursts associated with SEP events.
    \item We cross-verified the possible ESP events in our data set with the list by \cite{huttunen2009interplanetary} and find three ({\tt\string gsep$\_{167}$}, {\tt\string gsep$\_{216}$} and {\tt\string gsep$\_{223}$}) `confirmed' and two ({\tt\string gsep$\_{208}$} and {\tt\string gsep$\_{209}$}) `probable' ESP events. The remaining four events ({\tt\string gsep$\_{146}$}, {\tt\string gsep$\_{164}$}, {\tt\string gsep$\_{168}$} and {\tt\string gsep$\_{180}$}) are not reported in the former list. Nonetheless, we retain all these nine events in our catalog but indicate them as `ESP' events under the `$gsep\_{notes}$' column.
    \item Two large and weak SEP events each have no observational data on source solar eruptions.
    \item No flare has been reported on 1991-05-20 by NOAA. Hence, no source association is available for the weak SEP event observed that day.
\end{enumerate}

\section{Parent and SEP Event Properties} \label{sec:prop}
The initiation of an SEP event depends on either a flare or CME eruption or both. We observe two scenarios here: (1) There is at least a solar flare as a precursor to the SEP event; (2) There is a CME only as a precursor to the SEP event. The onset of a flare, its magnitude, and its location may influence the extent of the rise of proton fluxes and the respective fluences near Earth. Some SEPs connected with a flare and a CME typically show large fluxes near Earth even when they originate from the eastern side of the Sun. This could be partially due to the magnetic field lines getting stretched by the CMEs while the shock-fronts sweep the particles and accelerate them in the interplanetary region.

We find 41 SEP events with a parent flare that erupted before 12 to 24 hours of SEP event onset. In a general situation where there are two or more precursors, a flaring event is always initiated before the launch of the corresponding CME. Temporal measurements of CMEs such as launch time and their first appearance  are subject to errors. However, pertinent data were considered in our analysis based on observational evidence along with those reported in the CDAW catalog.

\begin{figure}[ht!]
\plotone{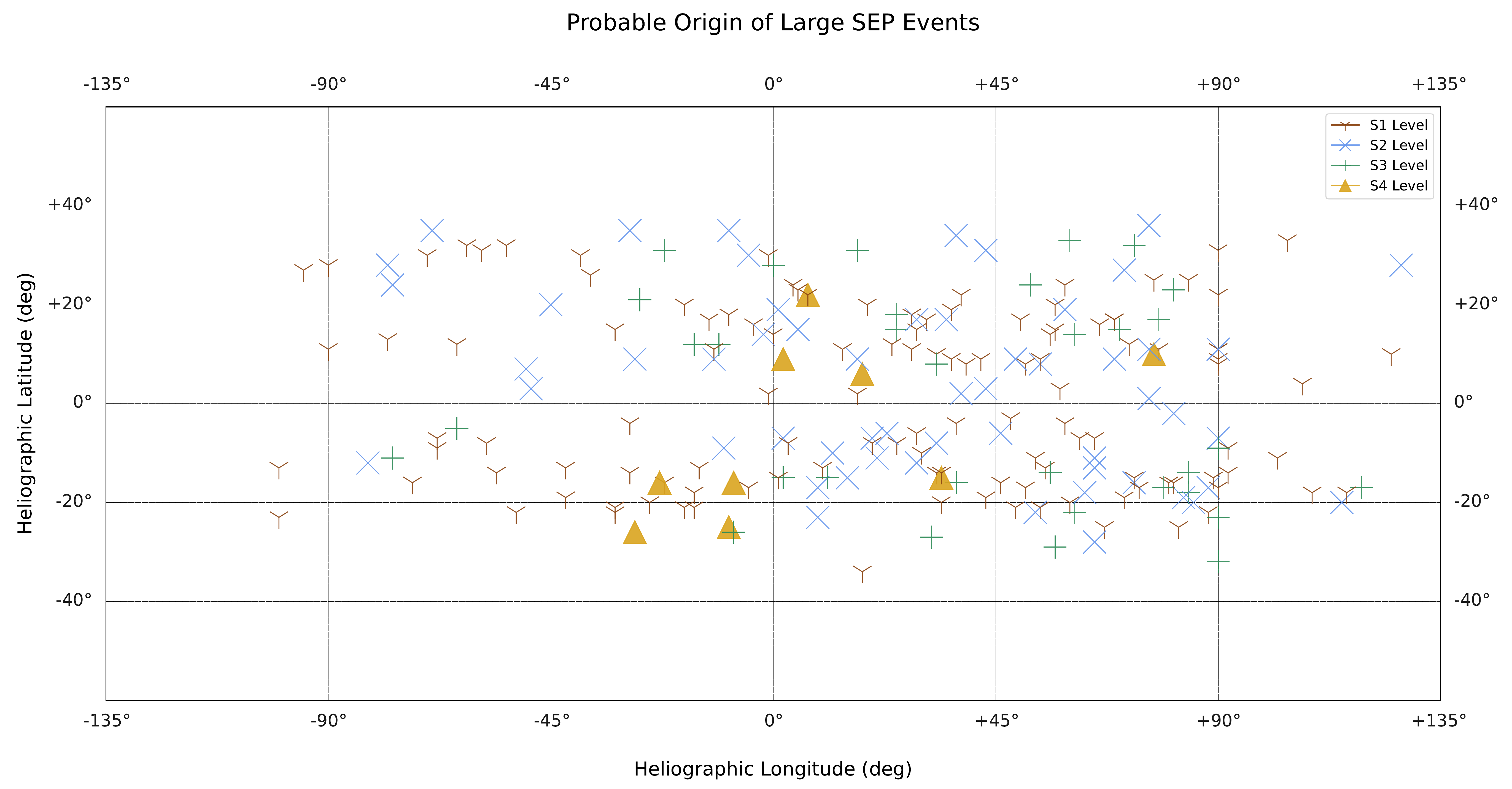}
\caption{Spatial variations of all the large SEP events in the GSEP dataset across the solar disk. The location values are plotted here as a function of the NOAA S-scale.
\label{fig:large_scatter}}
\end{figure}

\subsection{Solar Flares}
We utilize the NOAA-GOES flare list that provides data on the flare's onset, peak, end time, peak x-ray flux level (labeled as GOES class), coordinates of the flares and associated active regions based on \textit{in situ} soft X-ray (1–8Å) measurements by the GOES/XRS. Using the start and peak times of the flaring events reported in the NOAA-GOES flare list, we estimate the flare rise time, i.e., the number of minutes taken to reach maximum x-ray intensity. In the present work, erroneous flare associations were overcome based on the event temporal information; that flares must occur before an SEP event. In addition, the proton flux enhancements had to be observed during the soft X-ray emissions, and that the flare events were generally of long duration and intensity. Exceptions were cross-verified by visually checking the observational data.

\subsubsection{Active region locations}
Measurements of SEP events near Earth depend on the spatial region of source eruption on the Sun. Generally, it is understood that the eruptions at the western side of the Sun have a higher probability of energetic particles reaching near-Earth space due to the spiral structure of the interplanetary magnetic field lines, popularly known as the Parker spiral \citep{parker1965dynamical, reames1999particle}. Nonetheless, many SEP events have been detected arising from widely spread locations on the Sun. Figure \ref{fig:large_scatter} shows a cylindrical projection of flare coordinates that have given rise to large SEP events between 1986 and 2017. The location values are plotted here as a function of the NOAA S-scale to understand better the source location distribution and the corresponding SEP event peak intensity observed at Earth. It can be seen here that S1 and S2 level proton event origins appear concentrated on the western hemisphere but also are widely spread across the solar disk. A few S3-level SEP events originate from the eastern hemisphere and a few more from the western limb and regions beyond. Interestingly, three S4-level SEP events have been reported in SC22, six in SC23, and none in SC24. For these events, the minimum rise time is roughly 16 hours, with a mean of ~26 hours to reach peak fluxes. All nine events originated from magnetically well-connected regions between the Sun and Earth. However, the high intensity of S4 events and their large temporal extents here contradicts the commonly believed concept of SEP events from the western side displaying fast rise and short duration compared to eastern events.

\begin{figure}[ht!]
\plotone{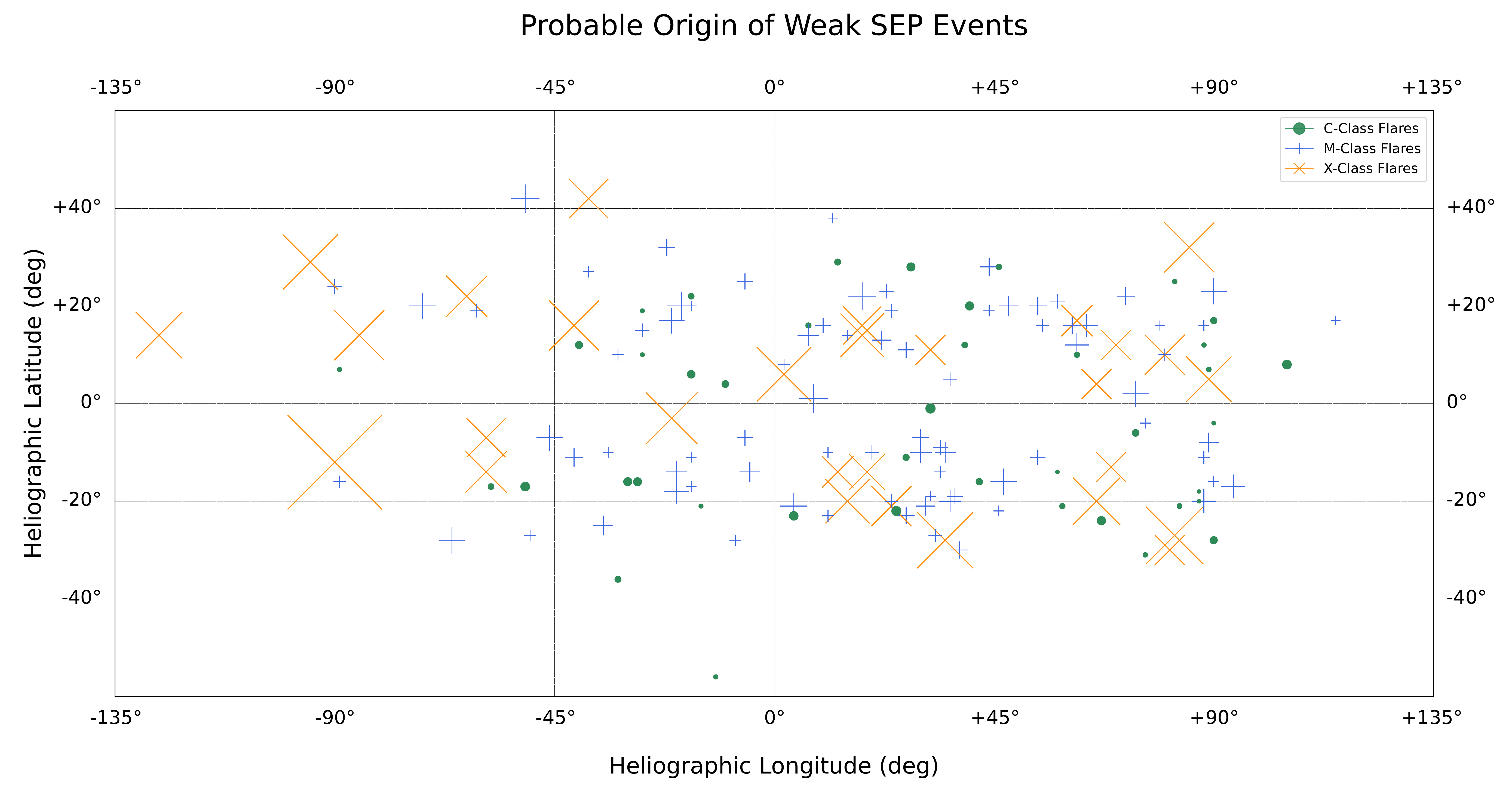}
\caption{Spatial variations of all the weak/sub-SEP events in the GSEP dataset across the solar disk. The location values are plotted here as a function of peak solar flare intensity. The X-ray flare measurements are obtained from the GOES missions.
\label{fig:sub_scatter}}
\end{figure}

For missing location values, we made efforts to gather information from additional sources such as other catalogs and images/movies from observations as mentioned earlier in section \ref{sec:data}. However, when there are inconsistencies and lack of logical conclusion we omit from including those erroneous information in our data set. For example, an SEP event on 1992-03-07 at 15:50 occurred over the eastern limb and is associated with a C5.0 flare with no report of an accurate flare onset. Another large SEP event on 1991-03-29 at 21:20 has no location information of a possible flare in the GOES flare list.

A weak SEP event on 2002-11-01 at 18:40 arises from the far side and lacks observational evidence to match with a source eruption. Figure \ref{fig:sub_scatter} shows a scatter plot of solar flare coordinates associated with weak SEP events between 1986 and 2017. Here, the location magnitudes are plotted as a function of the GOES flare class to understand the dependence of SEP peak fluxes at Earth on X-ray flare peak intensities. Note that the differences in sizes of the tick marks correspond to the GOES flare level. Apart from a few C-class flares in the far-eastern hemisphere, most of the flaring regions are present from -30 degrees to the far-western side. A similar trend can be noticed for M-class flares, except for those few beyond -45 degrees. However, X-class flare locations are widely spread across the Sun. Despite the significant flare intensities, associated proton flux levels remain $<$10 pfu near Earth. On comparing the two scatter plots in figures \ref{fig:large_scatter} and \ref{fig:sub_scatter}, it appears that there are several factors from the upper solar atmosphere to the corona, including the footpoints, that dominate the release and acceleration of SEPs from the Sun.

\begin{figure*}[ht!]
\gridline{\fig{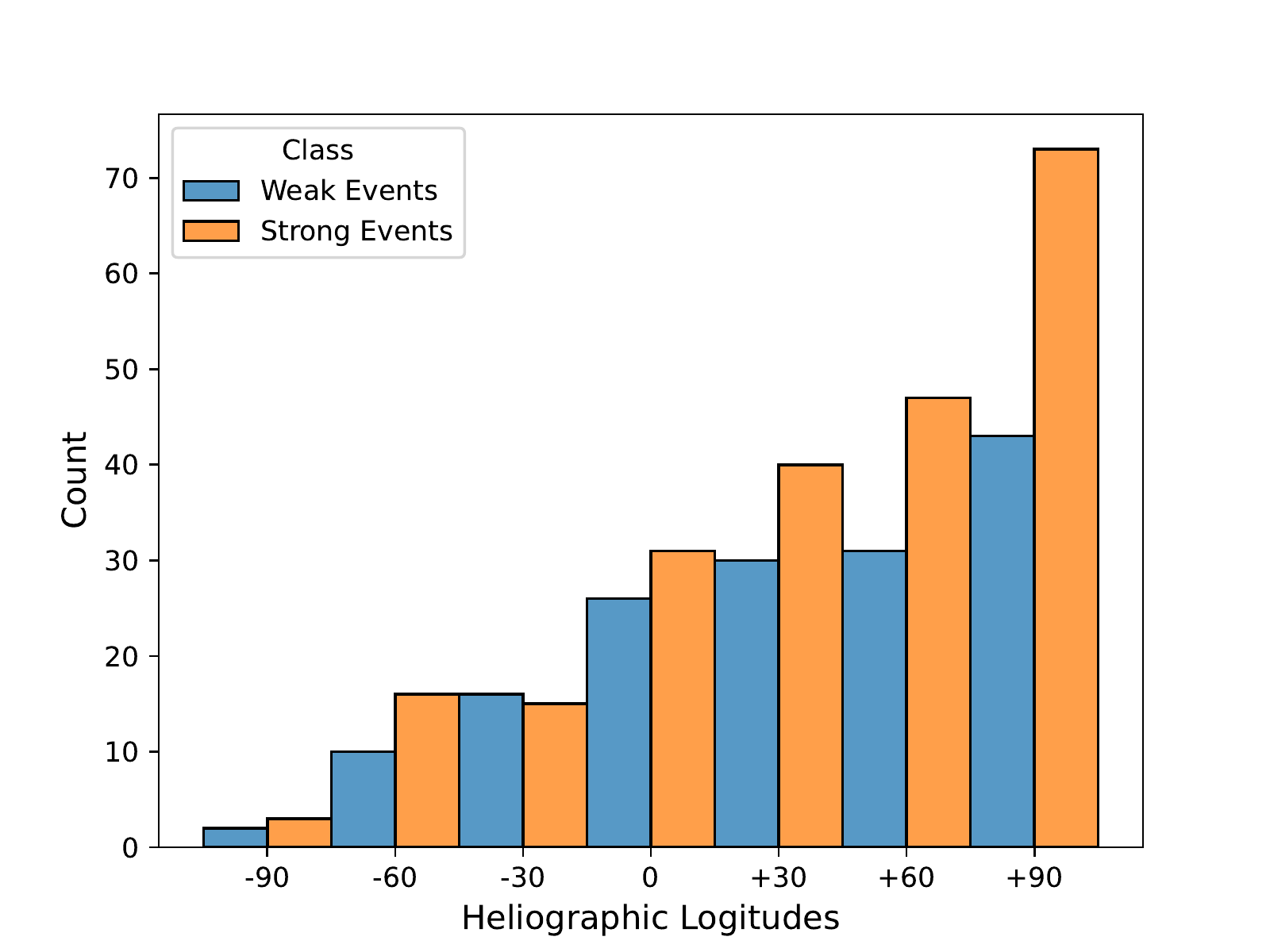}{0.45\textwidth}{(a)}
          \fig{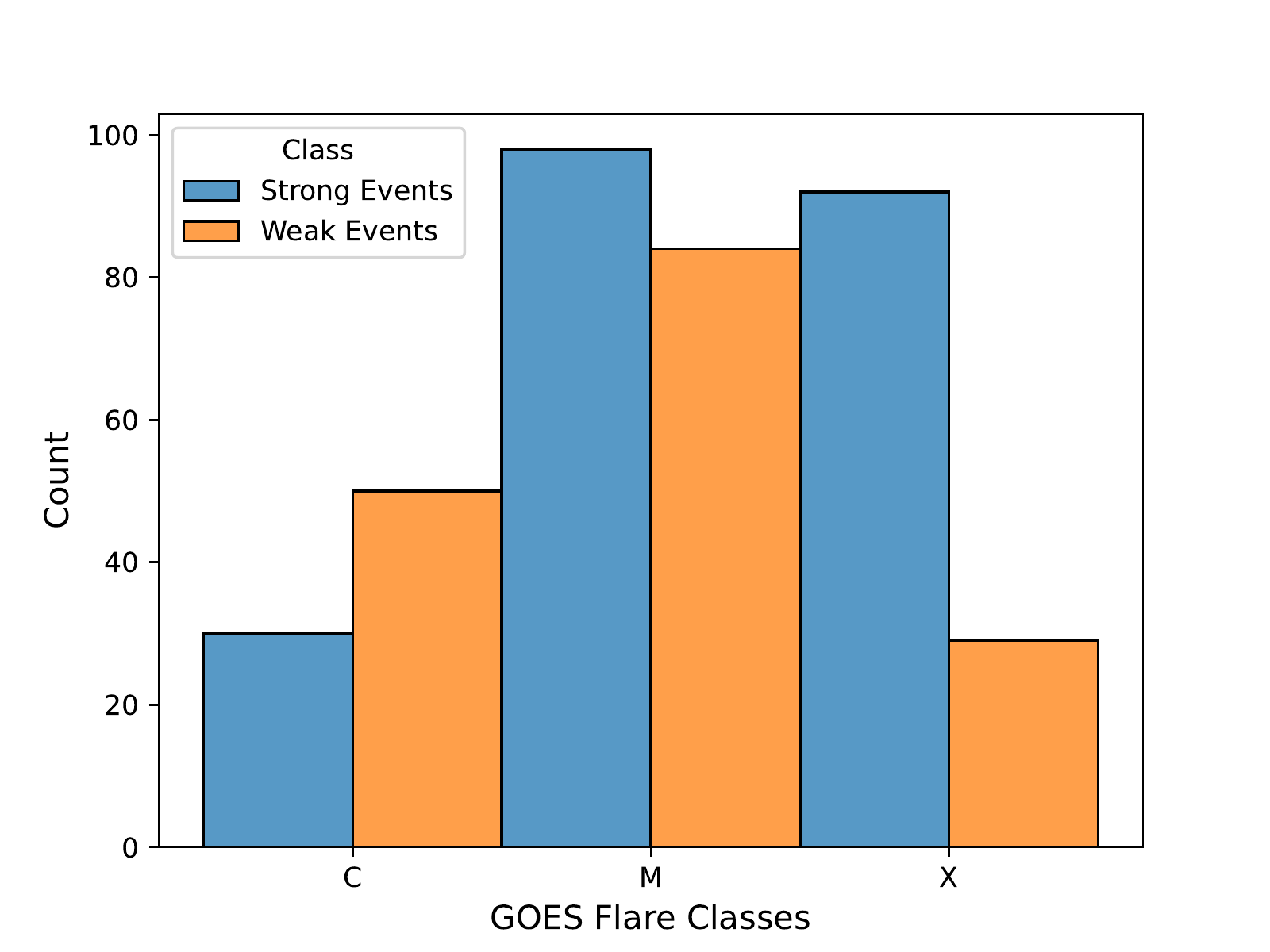}{0.45\textwidth}{(b)}
          }
\caption{Distribution of (a) longitudinal variation of solar flare locations that are associated with SEP events; (b) soft X-ray flare peak intensities based on GOES flare classification for flares associated with SEP events. The two classes of SEP events (strong and weak) are shown in the legend.
\label{fig:lon_var}}
\end{figure*}

Figure \ref{fig:lon_var}(a) shows the distribution of longitudes for 383 SEP-associated flaring events in our data set. The seven bins are calculated based on the number of events between a pair of longitudinal extents, namely, [$<$ -90], [$\geq$-90 to $<$ -60], [$\geq$-60 to $<$ -30], [$\geq$-30 to $<$ 0], [$\geq$0 to $<$ 30], [$\geq$30 to $<$ 60], [$\geq$60 to $<$ 90], and [$\geq$90]. There is an increase in the number of events as a function of flare longitude with strong preferences towards the western hemisphere.

\subsubsection{Flare Magnitudes}
Out of 433 events in our data set, there are 386 flare-associated SEP events. Here, 383 (220-large and 163-weak) events have GOES class information available. Of them, 121 (32\%) are associated with X-class flares, while 182 (48\%) and 80 (20\%) are associated with M and C- classes, respectively. Figure \ref{fig:lon_var}(b) shows the distribution of flare-associated SEP events where bin sizes correspond to the GOES flare classification. Weak SEP events show a closely normal distribution for source flare magnitudes. However, large events show a higher probability with M-class flares followed by X-s. Large SEP events have 92 - X, 98 – M, and 30 – C class flares associated, where the median flare class is M2.0. Weak SEP events have 29 - X, 84 – M, and 50 – C class flare associations.

\begin{figure*}[ht!]
\gridline{\fig{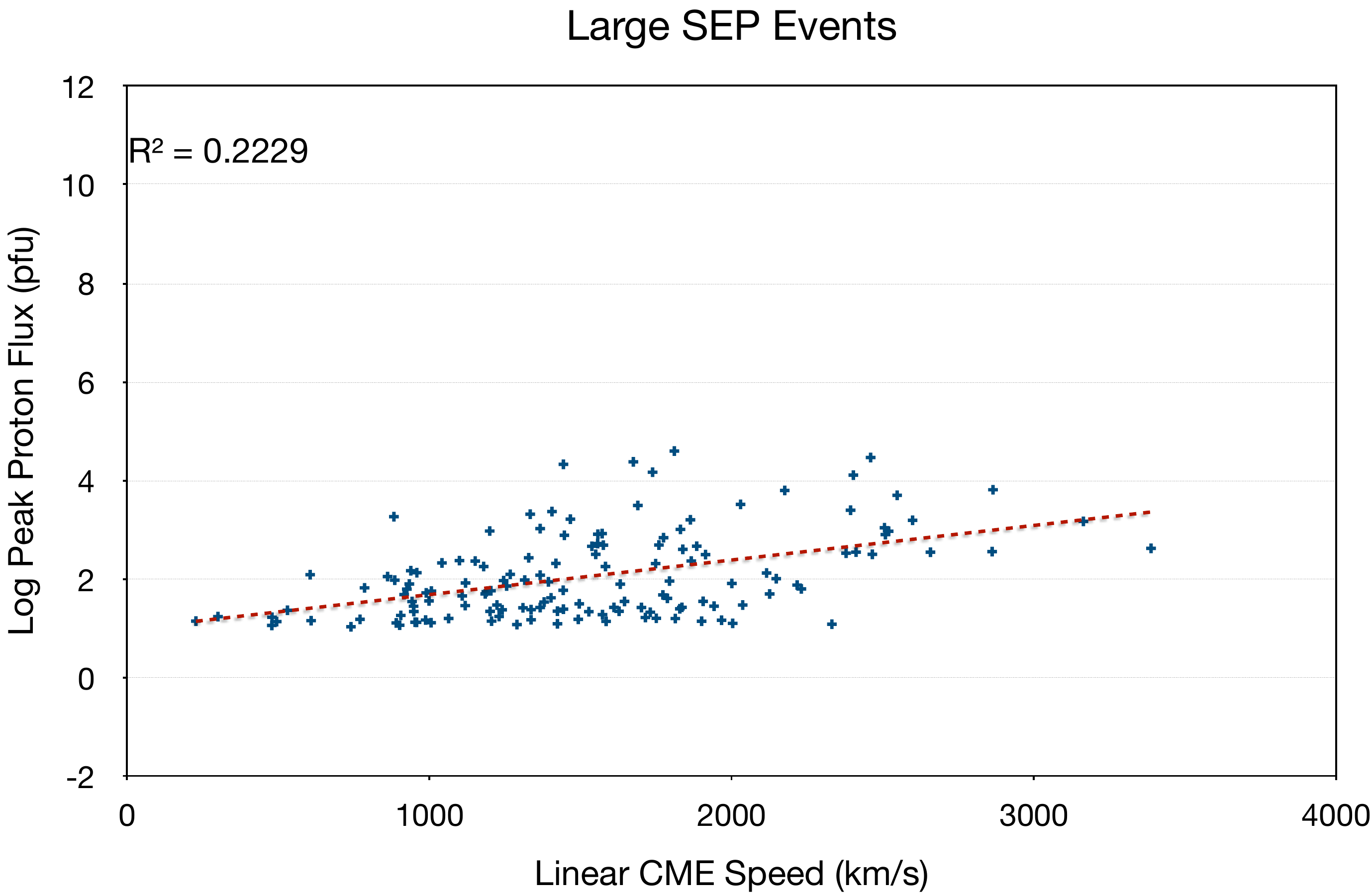}{0.45\textwidth}{(a)}
          \fig{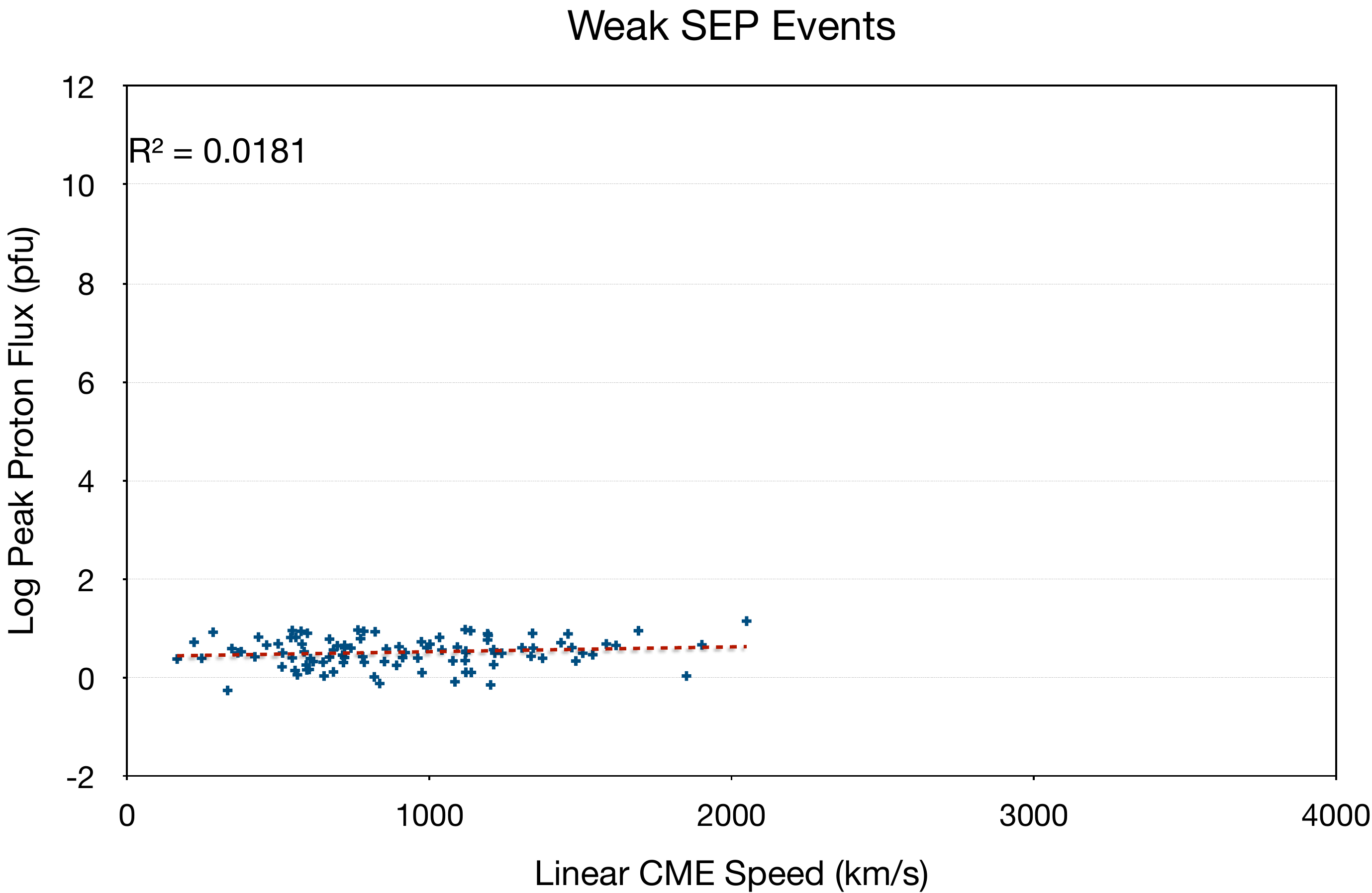}{0.45\textwidth}{(b)}
          }
\gridline{
          \fig{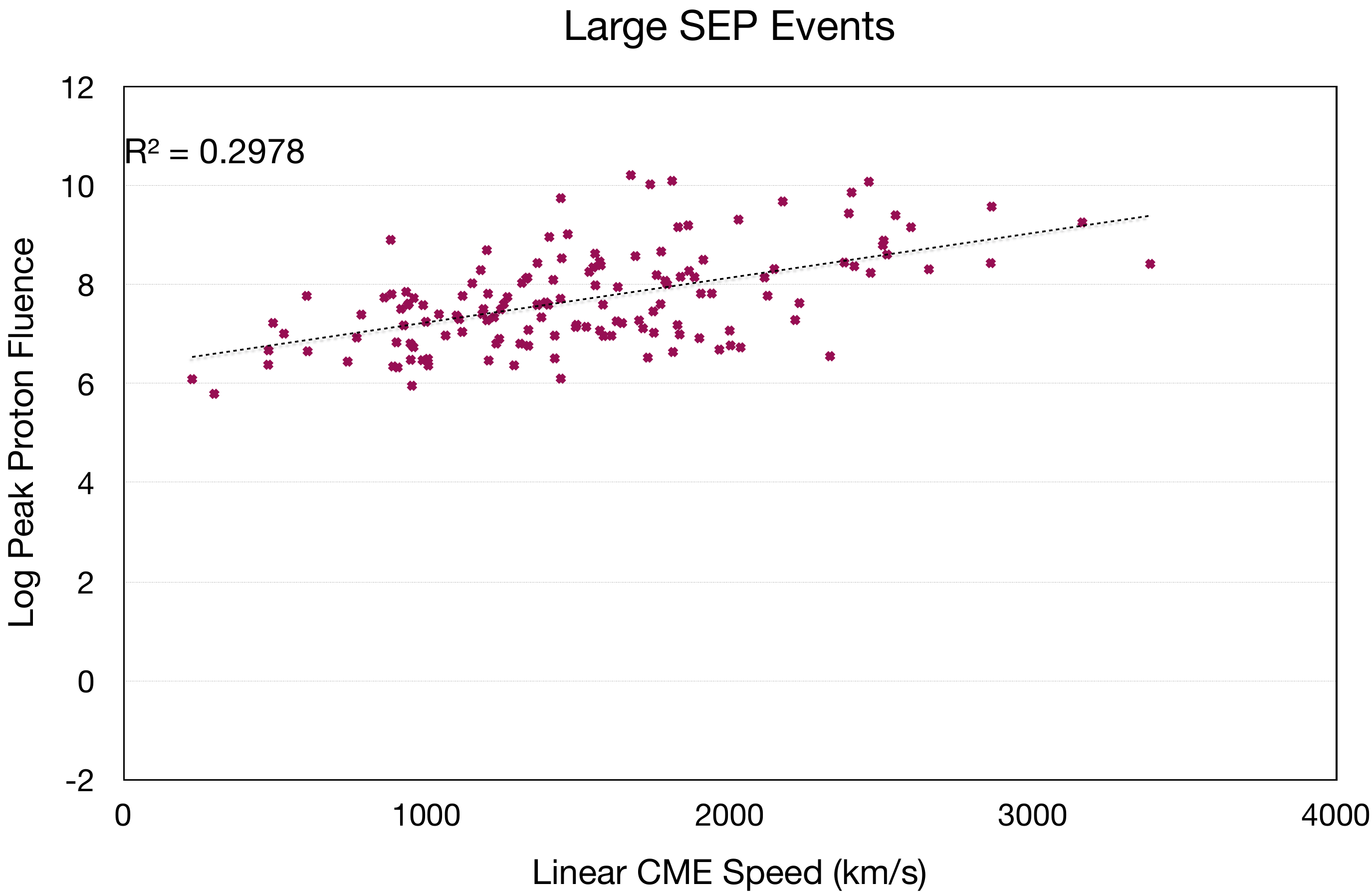}{0.45\textwidth}{(c)}
          \fig{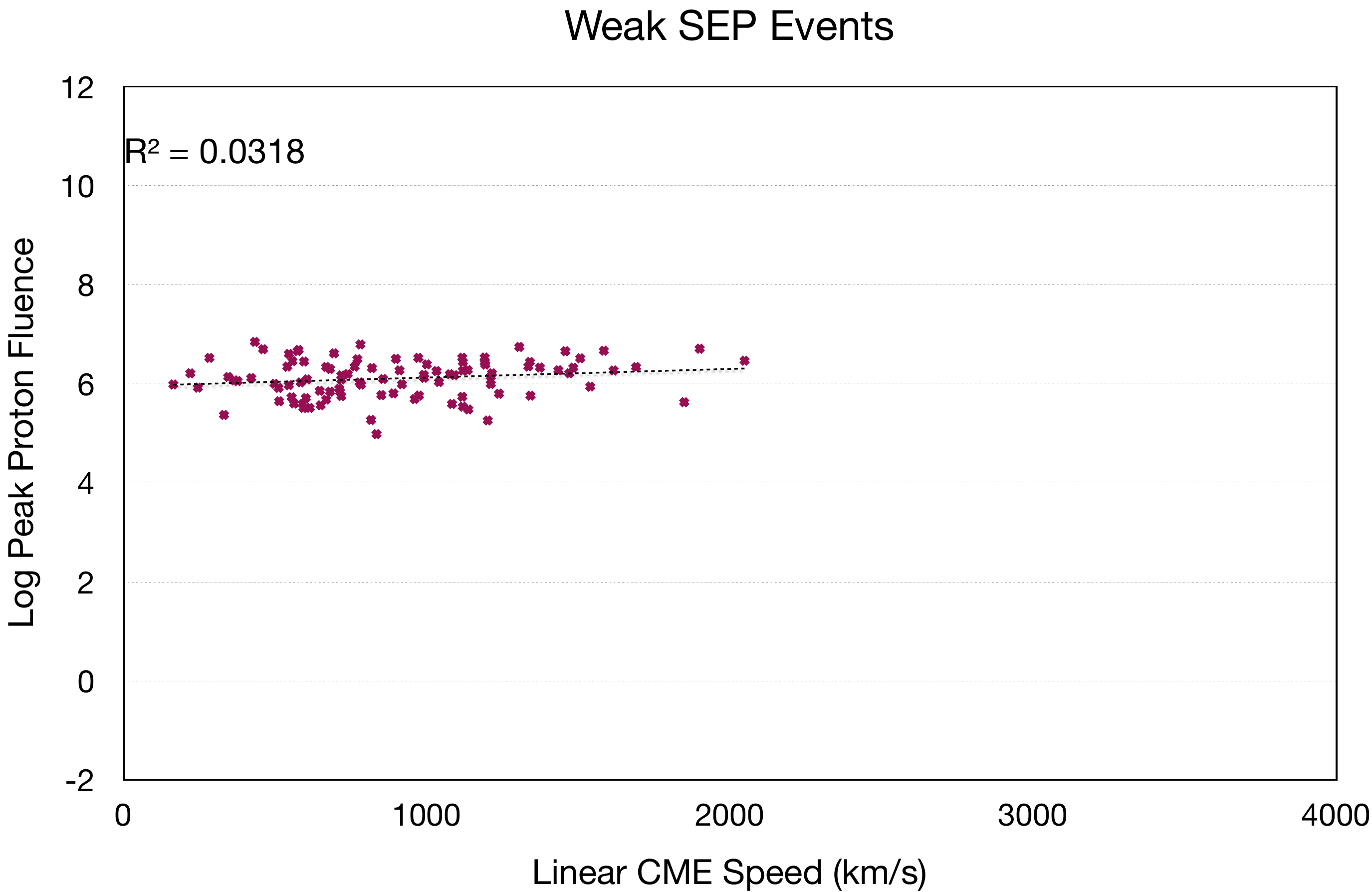}{0.45\textwidth}{(d)}
          }
\caption{Scatter plot of the log-transformed peak proton fluxes for (a) large and (b) weak SEP events. Similarly, we show a scatter plot of proton event fluences with respect to CME speeds for (c) large and (d) weak SEP events. A linear regression trend is fit to the data which shows the respective variance (R$^2$) of the distribution as an inset within the corresponding figure.
\label{fig:cme_scatter}}
\end{figure*}

\subsection{Coronal Mass Ejections}

The CDAW provided a list of CMEs as observed by the SOHO coronagraphs. Estimations and descriptions of various parameters, including CME speed and angular size have been made by \citet{gopalswamy2009soho} and are available from the official CDAW website. Since 1997, LASCO has observed that a CME preceded every SEP event except for eighteen cases, apart from five more events when there was sustained activity in 1998 \citep{yashiro2004catalog}.

\subsubsection{Source CME properties.}
In our data set, 247 SEP events have been identified with precursor CMEs. Here, 162 CMEs are halo with a median speed of 1366 km.s$^{-1}$. About 144 flares have been observed to associate with these halo CMEs, and 114 of them resulted in large SEP events as measured near Earth. SEP events associated with non-halo CMEs are 85 in number with a median speed of 767 km.s$^{-1}$. We obtain the CME speeds from the CDAW CME catalog publicly available online. In figure \ref{fig:cme_scatter}, we show the variance of logarithmic peak proton flux with respect to linear CME speed for (a) large and (b) weak SEP events in our data set. The distribution consists of a linear regression trend fit on the data and shows a 22$\%$ and 1$\%$ variance (R$^{2}$), respectively. In addition, logarithmically transformed proton fluences of large and weak events are shown in subsets (c) and (d), respectively. Here, the linear regression fit on the distribution shows $\approx$30$\%$ and 3$\%$ variance (R$^{2}$), respectively. As evident from the figures, CME speeds cannot be fully deterministic about SEP events, especially the weak class.

\begin{figure*}[ht!]
\gridline{\fig{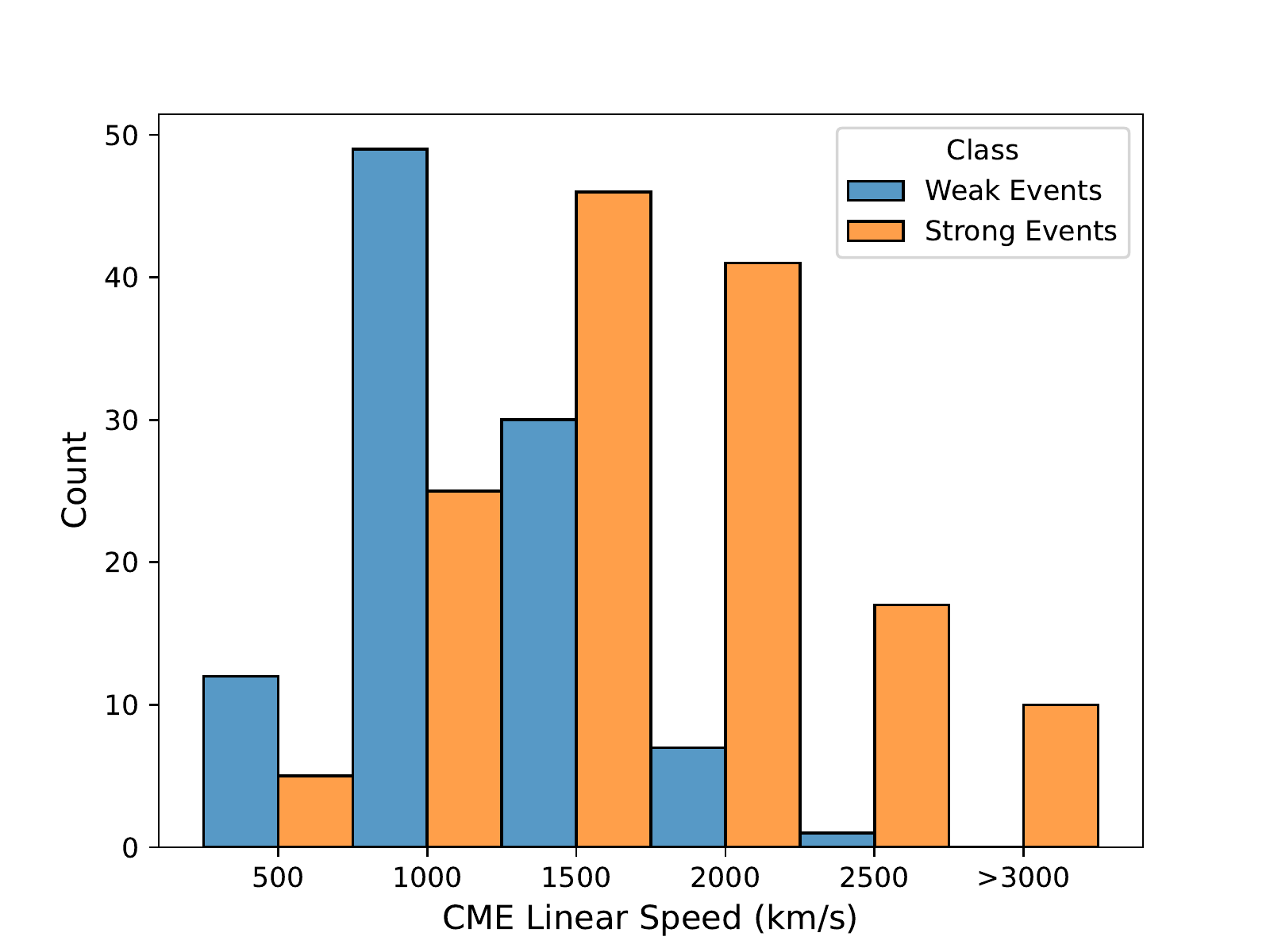}{0.45\textwidth}{(a)}
          \fig{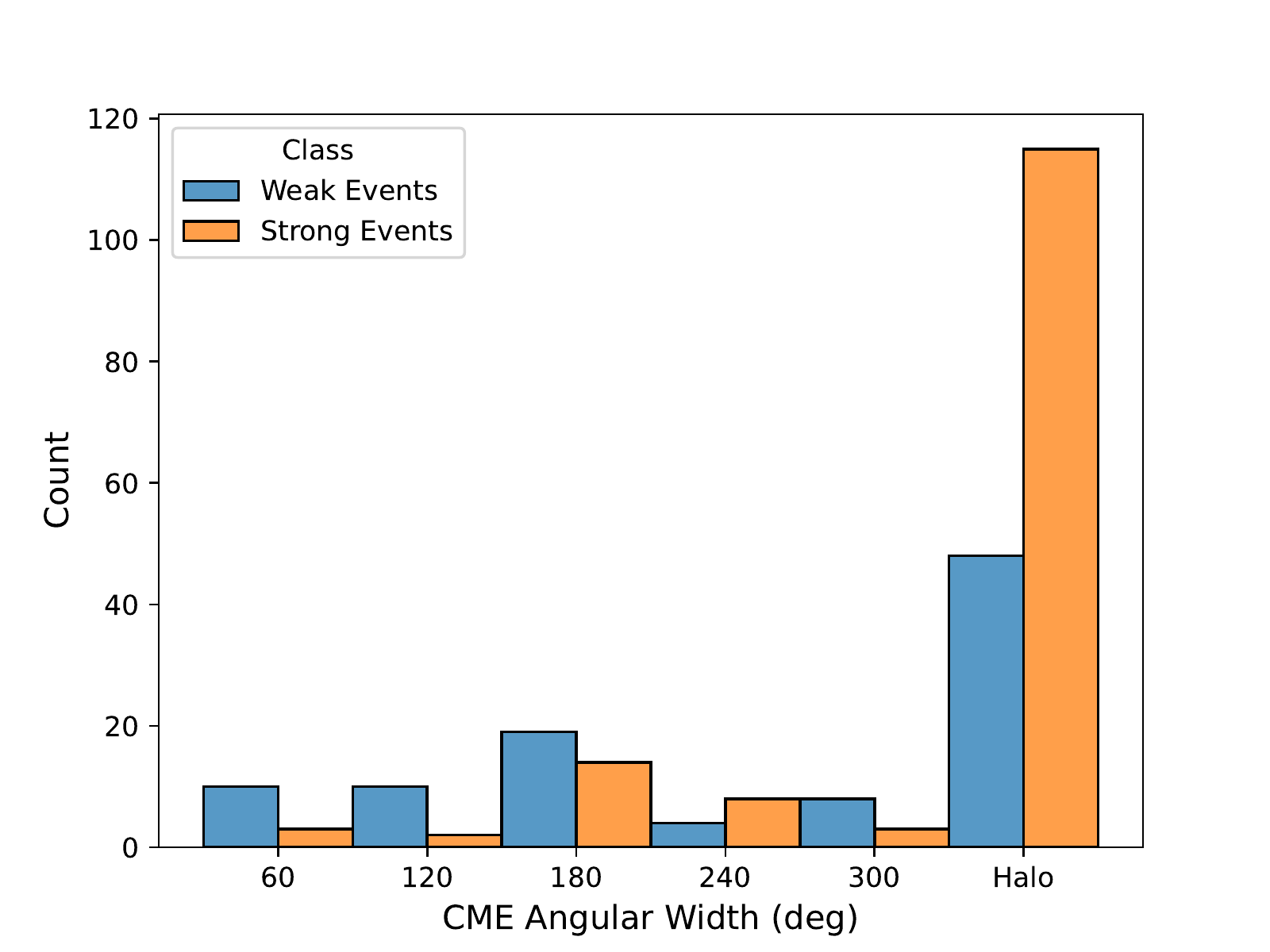}{0.45\textwidth}{(b)}
          }
\caption{Distribution of (a) linear speed and (b) angular width of CMEs that are associated with SEP events in the GSEP data set. The two classes of SEP events (strong and weak) are shown in the legend. It can be noticed that halo fast-CMEs appear to dominate the occurrence of large SEP events.
\label{fig:cme_hist}}
\end{figure*}

Shocks from halo-CMEs are dominant features as they stretch the interplanetary magnetic field, thereby expanding the horizon of field lines which in turn could cause the path of accelerating particles to stay away from Earth \citep{desai2016large}. Large SEP events ($\approx$70$\%$) in our data set are predominantly associated with halo fast-CMEs. Figure \ref{fig:cme_hist}(a) shows the distribution of the speed of CMEs that are associated with SEP events. Weak SEP events have a median (mean) speed of ~905 (835) km.s$^{-1}$, while large events have 1522 (1444) km.s$^{-1}$. It is worthwhile to mention here that the measurements of CME speeds suffer from projection effects. Nonetheless, several researchers have made efforts to infer as much accuracy as possible \citep{yashiro2004catalog, gopalswamy2009soho}. The number of SEP events depending on the associated CME widths is shown in figure \ref{fig:cme_hist}(b). The bins correspond to $<$60, 60-120, 120-180, 180-240, 240-300, and halo (360) measured in degrees. A median (mean) of 281$\degree$ (246$\degree$) is noted for weak events. In our study, we find most (66\%) of the CMEs associated with SEP events are halo in angular extents and fast with a median speed of 1200 km.s$^{-1}$. 

\subsection{Type-II radio bursts}

We consult and integrate the CDAW's type-II radio bursts catalog into our data set. The CDAW list constitutes radio data measurements from the WAVES experiment \citep{Bougeret1995SSRv} onboard the Wind spacecraft \citep{wind1995SSRv...71....5A}. Type-II radio bursts have been largely associated with fast-CMEs and are widely considered to be possible signatures of SEP events. It has also been observed that each type-II burst follows a CME eruption in the upper coronal region \citep{gs2019SunGe}. For a total of 270 SEP events in our data set (for SCs 23 and 24 only), 147 have been associated with a type-II radio burst. More than half of these bursts occur within three hours before the SEP event onset with the dominant feature of that day's spectral plot. In our data set, one radio signature is uncertain, and another is reported by CDAW to be caused by a CME-CME interaction. Three radio bursts occur prior to SEP events originating behind the eastern limb and 14 behind the western limb. In these scenarios, radio emissions may be occulted.  In the CDAW-SEP list, seven more SEP events are associated with type-II radio bursts in solar cycle 23 but are not reported in CDAW's Wind/WAVES type II burst catalog. Hence, we have omitted them from our data set. Note that we have not examined individual frequency radio profiles as that is a complex and time-consuming task beyond this work's scope.

\section{Results} \label{sec:results}

\subsection{SEP Event Temporal Properties.}
\begin{figure*}[ht!]
\gridline{\fig{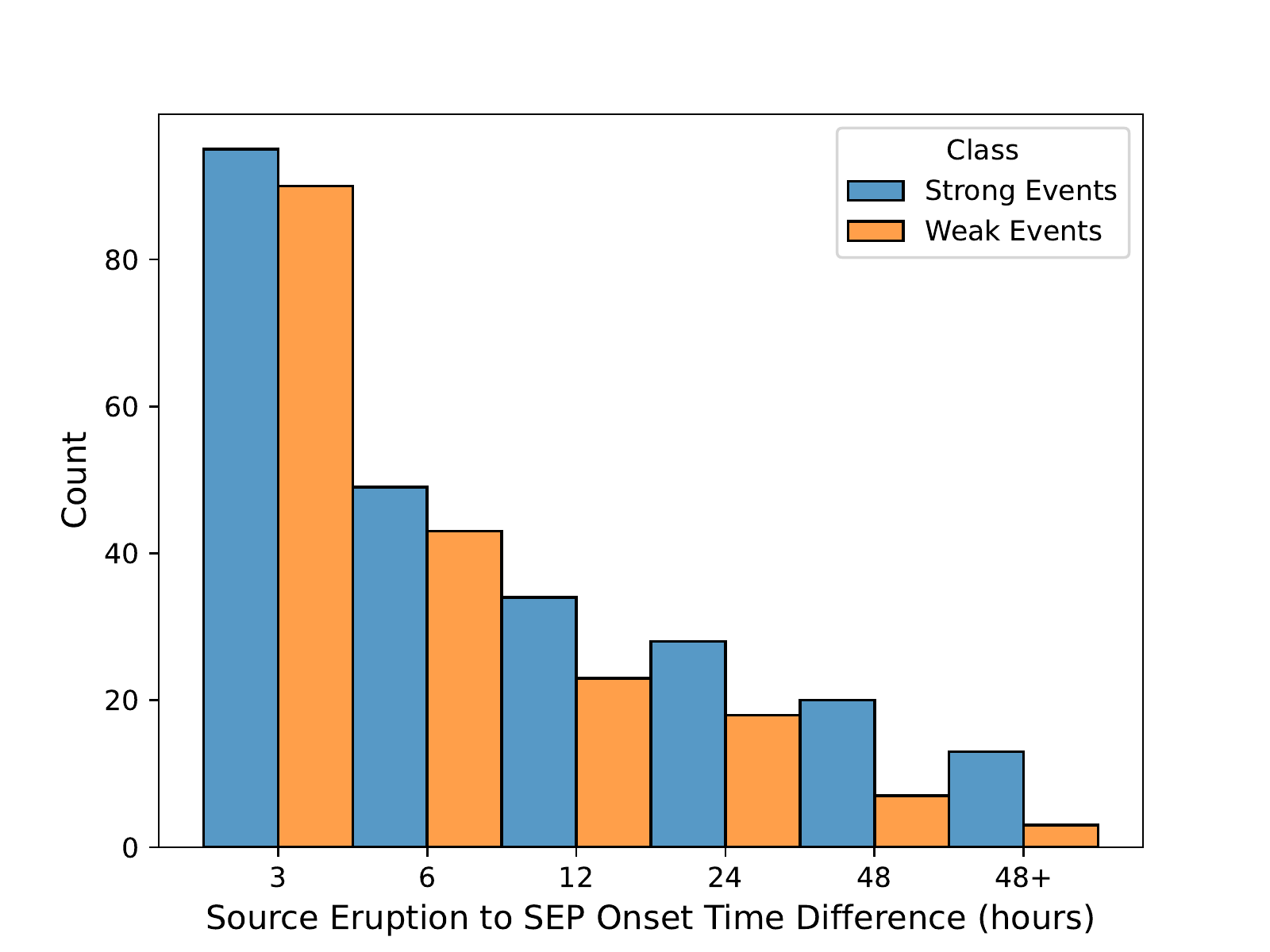}{0.45\textwidth}{(a)}
          \fig{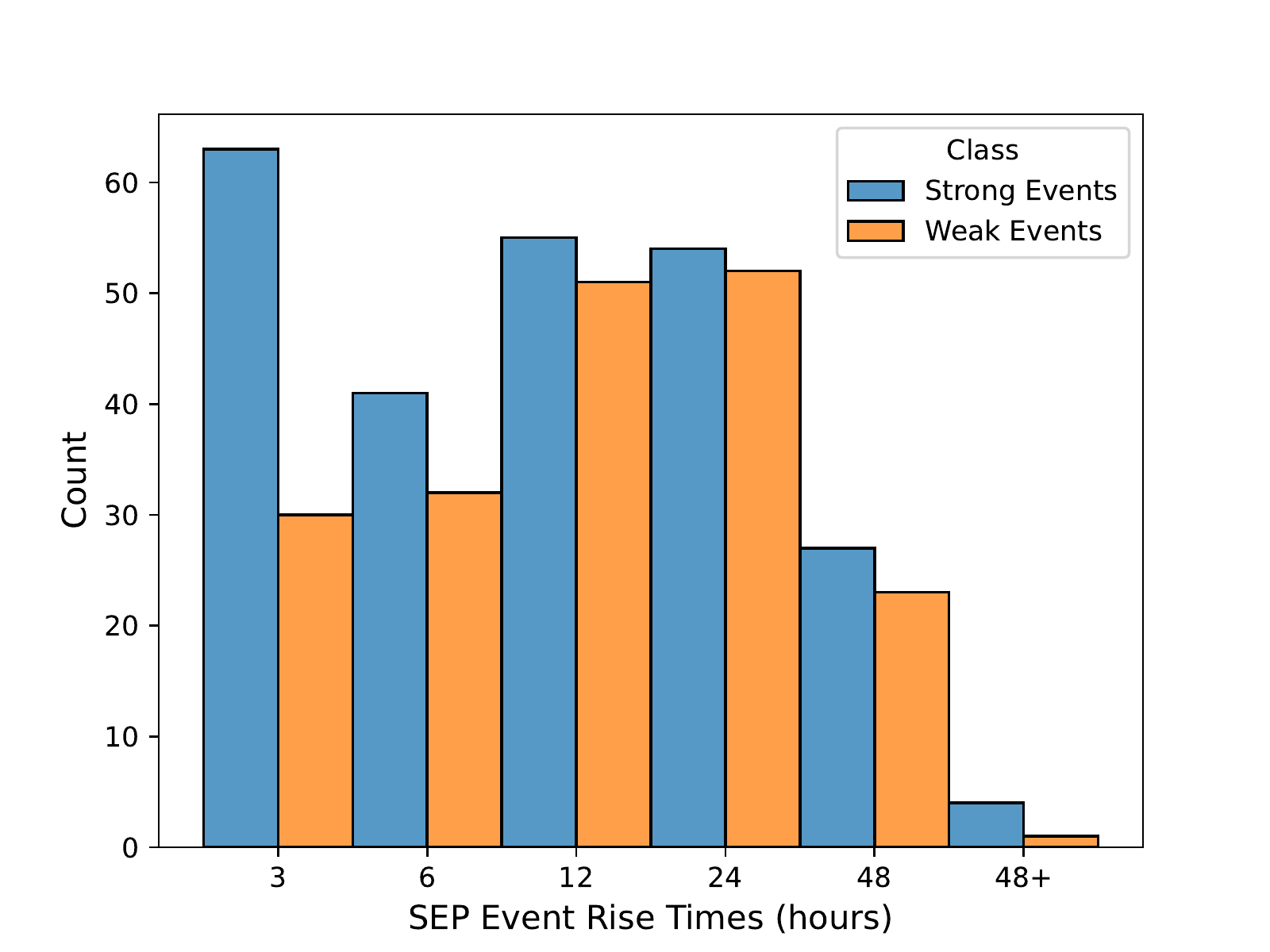}{0.45\textwidth}{(b)}
          }
\caption{Distribution of (a) time difference between source solar eruption (flare/CME) and SEP event onset. Up to 79$\%$ of SEP events arise from a precursor within 12 hours; (b) SEP event rise time, i.e., the time taken by the proton fluxes to reach peak intensities.
\label{fig:times}}
\end{figure*}

In addition to substantial variability in their elemental abundances, SEP events also exhibit variation in their time profiles. Hence, temporal variations between source solar eruption and the SEP onset become crucial in understanding the propagation of particles toward Earth. Figure \ref{fig:times}(a) shows a histogram of the time difference between a source (flare or CME) eruption and SEP event onset. As can be seen, 79\% (335 out of 423) of SEP events have a precursor eruption within 12 hours. 45 source eruptions occur between 12 to 24 hours, 28 between 24 to 48 hours, 13 between 48 to 72, and two SEP events have a source solar eruption over 72 hours before the SEP event onset. The median time delay between a halo-CME launch and SEP event onset is about 3.5 hours. At a median (average) of 26 (31) minutes after a CME is launched, a type-2 radio burst is initiated. The median (average) non-halo CME-RB time difference is 35 min (43 min).

Figure \ref{fig:times}(b) shows histograms of rise times (i.e., the time between the event onset and peak) in our data set, classified in bins of  0 - 3, 3 - 6, 6 - 12, 12 - 24, 24 - 48, and $>$48 hours. The overall median (average) rise time SEP events is 7.5 hours ($>$11.5 hours). For halo-CME-associated events, the median (average) rise time is $\sim$nine\ hours (13 hours). For non-halo associations, the median (average) SEP event rise time reduces to seven hours ($\sim$9.5\ hours).

\subsection{Source and SEP Events Correlation}
\begin{figure}[ht!]
\plotone{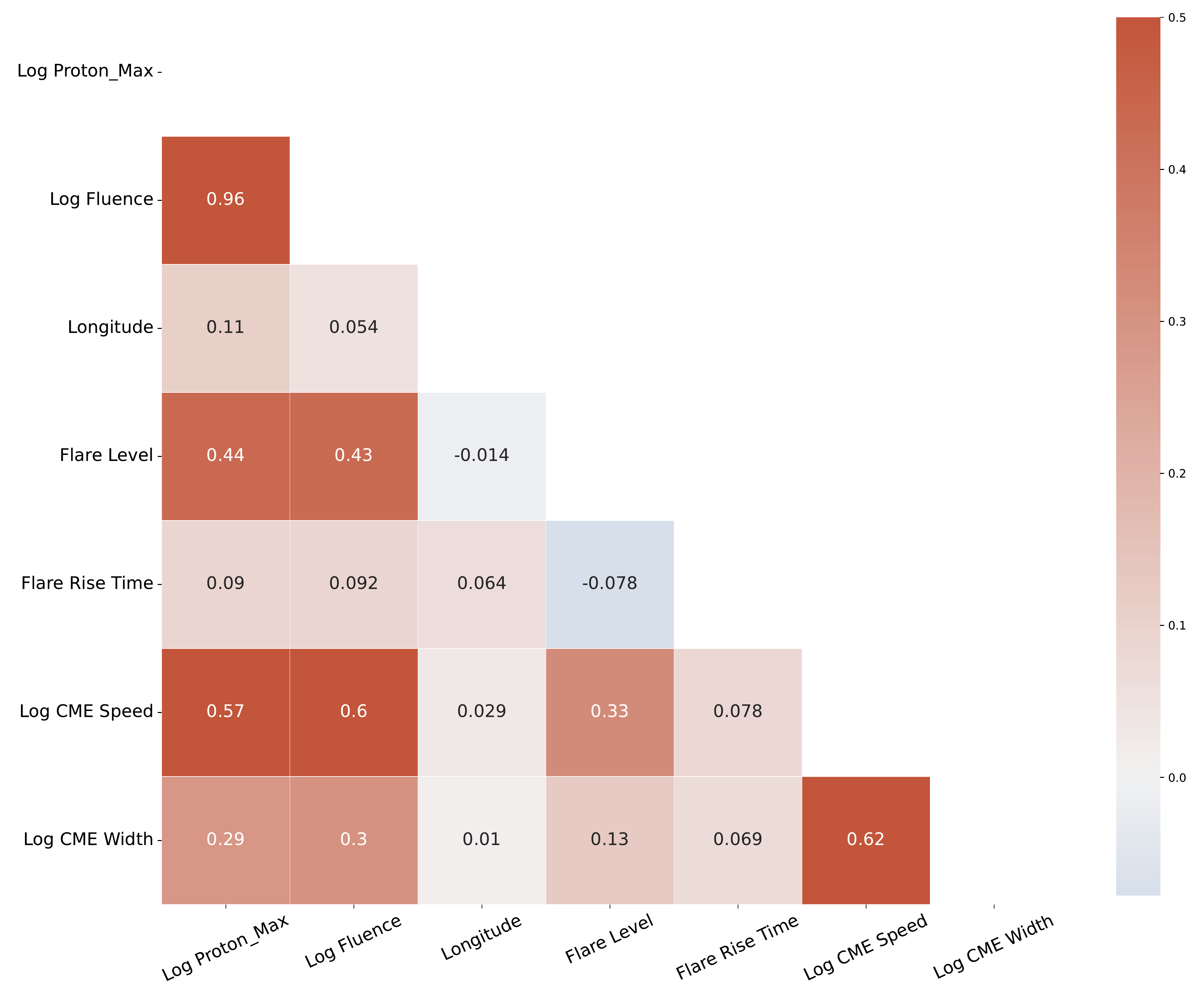}
\caption{Correlations coefficients of five parameters of source solar eruptions are obtained here for peak proton fluxes and fluences of SEP events. Logarithmic transformations of the CME and SEP variables are considered here. CME Speed shows a good correlation with SEP events, while flare intensities are moderately correlated.
\label{fig:corr}}
\end{figure}

We study the spatio-temporal dependence of SEP events on source eruptions by calculating the Pearson correlation coefficient of features in our data set. As discussed earlier, the peak of an SEP event, its fluence and rise time can be studied in relation to different source eruption features such as flare magnitude, flare rise time, corresponding active region longitude, CME speed and width. To examine the strengths between each of these variables, we construct a heatmap of the correlation coefficients as shown in figure \ref{fig:corr}. Notice that the logarithm of peak proton fluxes and proton event fluences have a good correlation with the logarithm of CME speed. There is a moderate correlation with flare level (X-ray magnitude) and a low correlation with the logarithm of CME width. There is also a good correlation between the logs of CME speed and width. In addition, the flare magnitude and log of CME speed show a low correlation coefficient. But features such as flare rise time and flaring location (longitudes) are poorly correlated with the rest of the variables. A possible scenario for the correlations of CME parameters not to be strong could be due to interacting CMEs. Interactions occur when a fast CME erupts immediately right behind a relatively slower CME or occurs too close in time to preceding events. Such back-to-back interacting CMEs allow an additional flux of particles to be accelerated in the interplanetary medium. However, analyzing these cases one by one is highly complex and is out of the scope of the current work.

\subsection{Supervised Machine Learning Implementation}

In machine learning (ML), one of the preprocessing techniques to improve the model performance is the reduction of dimensionality of the data set by identifying and choosing the most essential features \citep{khalid, ghareb_hybrid_2016}. Selecting highly relevant features from a data set can then be used as a core subset in training  models \citep{wei_novel_2020}. This step has multiple purposes, such as improving the performance of the models by eliminating unimportant features, providing robust predictors thereby reducing computational costs, and offering better interpretability to the underlying physical process that generated the data model \citep{guyon2003introduction}. In this section, we present our efforts in examining feature importance in our data set. All our computational experiments are performed using the Python programming language.

In the previous section, we examined the correlations of solar source parameters with SEP events and obtained some insights on possible features that can be relevant to SEP predictions. However, another way to confirm the existing relationships of those relevant features can be implemented using tree-based ML models by extracting information `gain' on each feature \citep{janitza, prasetiyowati}. For this purpose, we utilize random forest (RF; \citet{breiman_random_2001}), and extreme gradient boosting (XGBoost; \citet{XGB}) classifiers. These two models are popular tree-based learners that are extensively used in many areas of active research \citep{pal_random_2005, fawagreh_random_2014, sarica_random_2017, tyralis_brief_2019, korsos_testing_2021, can_comprehensive_2021, osman_extreme_2021, lavasa2021assessing}.

RF is a collection of decision trees on subsets of the data where the average of the prediction from each tree is obtained and based on a majority vote, the final output is predicted. RF follows the bagging technique (parallel building) to improve its performance. XGBoost builds on the gradient boosting model, that is, implementing the gradient decent algorithm in parallel and improving its performance in each step.

\subsubsection{Model Generation}
We consider large SEP events as a `positive' class and weak events as a `negative' thereby designing the problem as a binary classification task in this work. The six features considered are magnitude, rise times and location coordinates of flares, log-transformed speeds and widths of CMEs, and the time difference between source eruption and SEP event onset. Because we want to consider both the SF and CME parameters to classify the SEP events, we drop all rows with any null values in our desired feature columns. Doing that, the size of the data set has been undersampled to 146 instances preserving less number of (37) weak SEP events. We standardize our features by implementing the {\tt\string StandardScaler} module from scikit-learn. We hold 40$\%$ of the data as the test set to evaluate the performance of the models.

Traditionally, a third validation set is partitioned from the data to tune the hyperparameters and evaluate the model. However, this results in further reduction of sample size to help the model learn the pattern in the data. Hence, we implement a procedure called k-fold cross-validation (CV) where the training data is split into k number of smaller sets or folds \citep{Mosteller:68, stone1977asymptotics}. The model is first trained on k-1 folds and the resulting model is validated on the held-out fold. The overall performance of k-fold CV is evaluated as the average of the values computed in each step. In our work, we choose k = 2.

We determine the best parameters of the models based on grid search hyperparameter optimization techniques. For our tree-based models, we experimented with the maximum depth of the tree and the number of estimators. The best-case scenario for both the models resulted in 100 estimators and a maximum depth of four. We handled the class imbalance ratio between positive and negative instances by assigning appropriate parameters to automatically adjust and balance the weights based on class frequencies. 

\subsubsection{Feature Importance}
Statistical findings from earlier works (such as \cite{cane2010study, trottet2015statistical, papaioannou2016solar, anastasiadis2019solar}) emphasize the importance of SF and CME features leading to SEP events. In addition, recent work by \cite{lavasa2021assessing} also implemented model-based feature importance and finds the CME speed, width and soft x-ray fluence as prominent features relevant to SEP event prediction.

\begin{figure*}[ht!]
\gridline{
          \fig{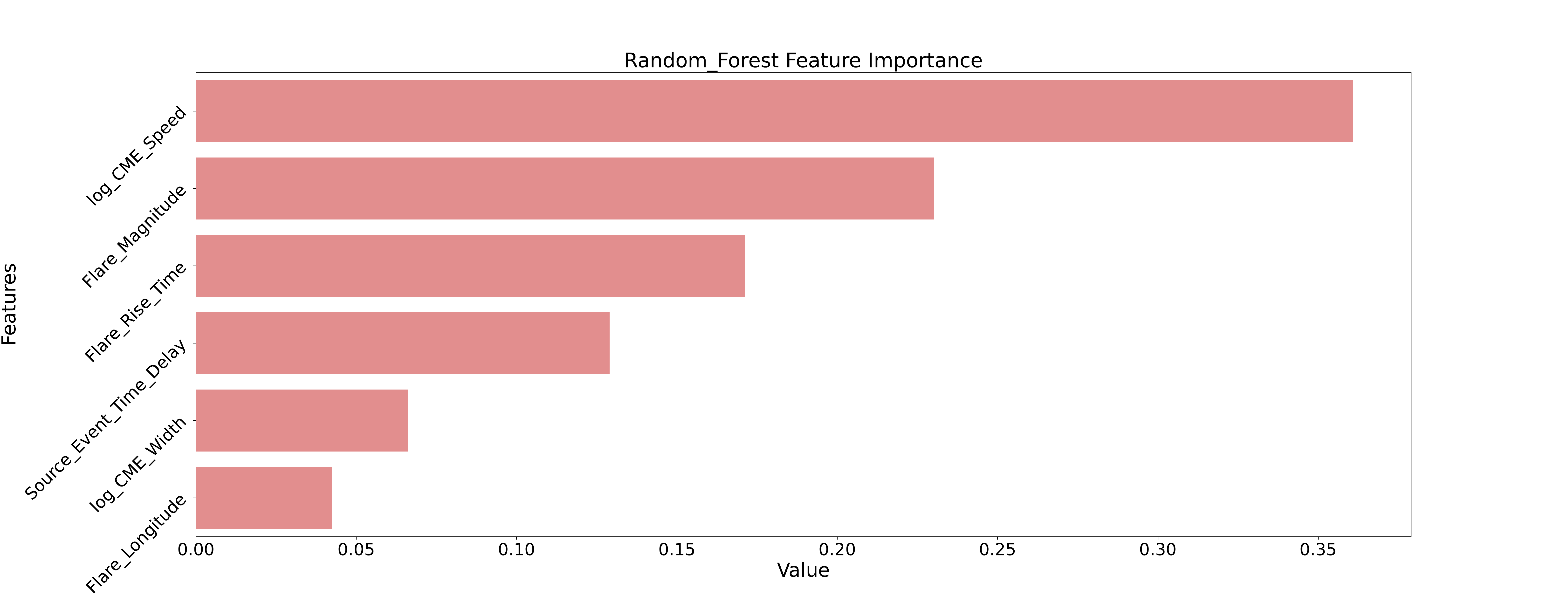}{0.95\textwidth}{(a)}
          }
\gridline{
          \fig{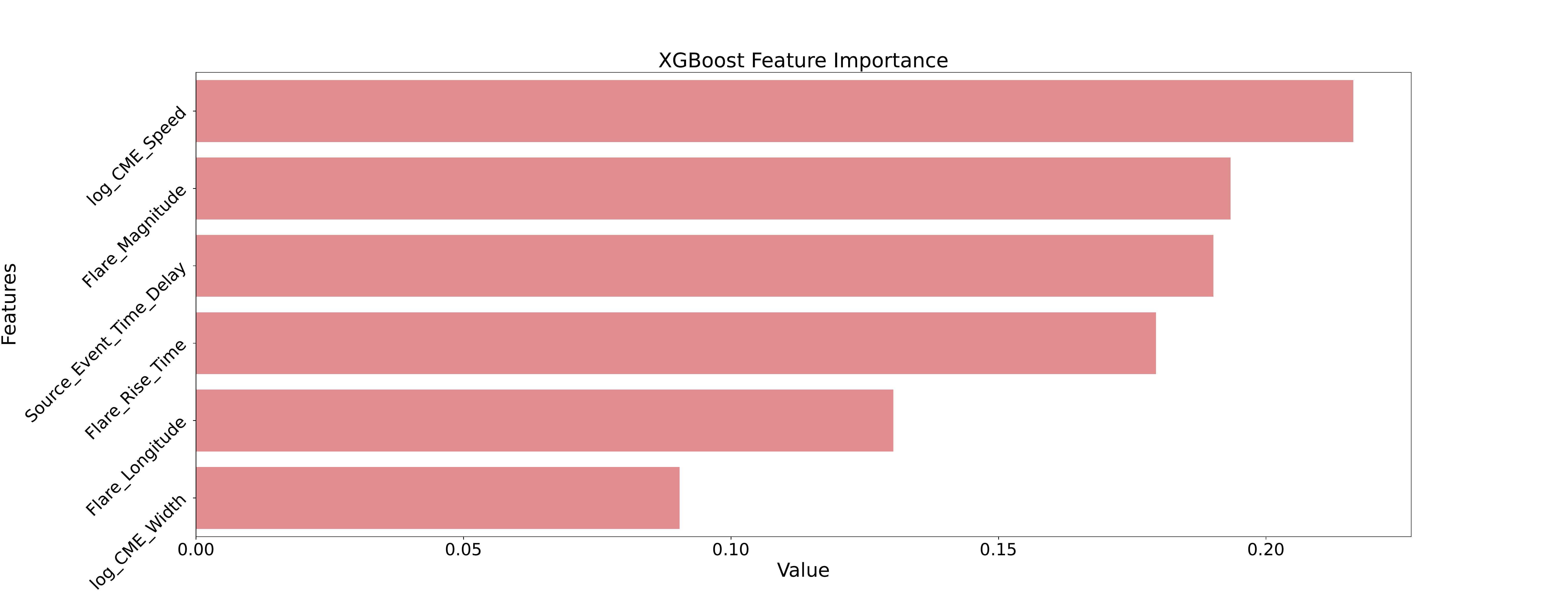}{0.95\textwidth}{(b)}
          }
\caption{Ranked results of feature importance in our data set based on (a) random forest and (b) XGBoost models. Both models build on both flare and CME parameters where logarithm of CME speed and the flare magnitude are the top two highly significant features.
\label{fig:fi}}
\end{figure*}

Both the tree-based models used in our study have a built-in {$feature\_importance$} attribute provided by the scikit-learn library\footnote{\url{https://scikit-learn.org/stable/auto_examples/ensemble/plot_forest_importances.html}} \citep{scikit-learn} that can be used to understand the splitting criterion within our models. Each feature is assigned a score that indicates how informative it is towards predicting our target variable. Therefore, higher-scoring features are more important. Using this technique, we extracted feature importance from our data set, and show the rank of our models in figure \ref{fig:fi}. Here, we can see that there is a close match in the results between models. Both RF and XGBoost have used ``log\_CME\_speed'' and ``flare magnitude'' as the top two important features. Nonetheless, the slight difference in ranking of the rest of features arises from the fact that RF randomly selects the features for each tree while sampling the data set. After that, the importance of a feature is estimated as a total reduction in the criterion due to that feature \citep{scikit-learn}. On the other hand, XGBoost uses features that have higher correlation between them to split the trees. The chosen top feature then remains constant throughout the model. Nonetheless, we conclude from our feature importance results that both the SF and CME parameters (CME speed and flare intensity, respectively) are indispensible in understanding and predicting the SEP event occurrence. 

\begin{deluxetable*}{llcccccc}[ht!]
\tablenum{2}
\tablecaption{List of recently developed ML-based models to predict SEP events. \label{tab:seps}}
\tablewidth{0pt}
\tablehead{
\textbf{Source}   & \textbf{Period}   & \multicolumn{5}{c}{\textbf{Solar Features}}     \\
                  &                   & \textbf{Active Region} &\textbf{X-Ray/Flare} & \textbf{CME} & \textbf{Radio Bursts} & \textbf{Protons} \\
}
\startdata
\cite{boubrahimi2017prediction} & 1997 - 2013 & N & Y & N & N & Y \\
\cite{engell2017sprints} & 1986 - 2018 & N & Y & N & N & Y \\
\cite{lavasa2021assessing} & 1988 - 2013 & N & Y  & Y & N & Y \\
\cite{aminalragia2021solar} & 1988 - 2013 & N & Y & N & N & Y \\
\cite{stumpo2021open} & 1995 - 2017  & N & Y & N & Y & Y \\
\cite{sadykov2021prediction} & 2010 - 2019   & Y & Y & N & Y & Y \\
\cite{kasapis2022interpretable} & 1996 - 2010 & Y & Y & N & N & N \\
\enddata
\tablecomments{Binary values in columns 3 to 7 indicate whether the corresponding solar event feature was considered in the corrresponding study. That is, `Y' means yes, and `N' otherwise. }
\end{deluxetable*}

\subsubsection{SEP Event Classification}
In table \ref{tab:seps}, we list all the available studies on SEP event predictions based on ML models that have been developed in the recent past. Each study here implements a variety of popular models such as K-nearest neighbors (KNN), logistic regression (LR), decision trees (DT), extremely randomized trees (XT), support vector machines (SVM), and neural networks (NN) to obtain promising results. The models in these studies have been developed focusing on a combination of  variety of solar parameters such as AR features to fluxes of radio, X-rays and protons exhibiting high correlations between each other. In addition, the time period considered in these studies depend on the availability of their desired data.

In our work, we approach the SEP event prediction problem in a different perspective based on the GSEP data set. As there is no one-to-one correspondence between the data and the methods implemented, we do not compare our results with earlier studies. Nonetheless, we present below the results of our best models to classify the SEP events as ``large" and ``weak" in our data set. In addition to RF and XGBoost, we explore the performances of LR \citep{lr_cox} and SVM \citep{cortesSVM} classifiers in our work. We emphasize that the best-performing hyperparameters were chosen for our models to obtain robust results.

\subsubsection{Evaluation}
 A 2 $\times$ 2 contingency table is implemented in our work for a binary classification task (large/weak SEP events). This table constitutes the following scores: true positives (TP), true negatives (TN), false positives (FP), and false negatives (FN). Here, TP indicates the number of correctly predicted large SEP events (positive class) by a model while TN represents the number of rightly predicted weak SEP events (negative class). FP corresponds to the number of weak events predicted as large (false alarms) while FN corresponds to the number of large events predicted as weak (misses). Subsequently, the aim of our best model should be to reduce incorrect results represented by both FP and FN.

We use F1 score, Matthews correlation coefficient (MCC) \citep{Matthews1975ComparisonOT}, true skill statistics (TSS) \citep{woodcock_evaluation_1976, daan1985forecast} and Heidke skill score (HSS) \citep{heidke_berechnung_1926} as our evaluation metrics to estimate the performance of the models on training and test sets. Focusing on the importance of positive classes, we consider the F1 score that can be estimated as the harmonic mean of `precision' (TP/(TP + FP)) and `recall' (TP/(TP + FN)) as shown in equation 1.

Because we have an imbalanced data set, we prefer the following metrics that take into account the variation in the sizes of the two classes of target variable. Using the confusion matrix, TSS is defined to account for the false positive rate as shown in equation 2. However, accounting the true negatives to assess the performance for a binary class problem is important in our context. Hence, we also choose MCC and HSS as defined in equations (3) and (4), respectively.

\begin{equation}
F{_1} score = 2 \times \frac{(Precision \times Recall)}{(Precision + Recall)}
\end{equation}

\begin{equation}
TSS = \frac{TP}{TP + FN} - \frac{FP}{FP + TN}
\end{equation}

\begin{equation}
MCC = \frac{(TP \times TN) - (FP \times FN)}{\sqrt{(TP + FP) \times (TP + FN) \times (TP + FP) \times (TN + FN)}}
\end{equation}

\begin{equation}
HSS = \frac{2 \times ((TP \times TN) - (FP \times FN))}{(TP + FN) \times (TN + FN) + (FP + TN) \times (FP + TP)}
\end{equation}

The results of our best-performing models on the test set are presented in table \ref{tab:metrics}. The models are ranked based on the F1 score where XGBoost performs higher than the rest of the models. On the other hand, LR and SVM models perform better in terms of TSS, MCC, and HSS compared to the trees-based ones. These results give us hope for improvement based on further feature extraction and experimentation with a variety of ML models.

\begin{deluxetable*}{lcccc}[ht!]
\tablenum{3}
\tablecaption{Performance of supervised classifiers on the GSEP data set. \label{tab:metrics}}
\tablewidth{0pt}
\tablehead{
\textbf{Model} & \textbf{F1$\_$score} & \textbf{TSS} & \textbf{MCC} & \textbf{HSS} \\
}
\startdata
XGBoost     &   0.85    &   0.41   &   0.43    &    0.43    \\
RF      &   0.80    &   0.40    &   0.38    &   0.38    \\
LR      &   0.77    &   0.58   &   0.52     &   0.46    \\
SVM     &   0.71    &   0.58   &   0.52     &   0.46    \\
\enddata
\tablecomments{Class metrics are presented here for the best models implemented on the test set. \newline
Model names are: XGBoost - Extreme Gradient Boosting; RF - Random Forest; LR - Logistic Regression; SVM - Support Vector Machine. \newline
Metric names are: TSS - True skill statistics; MCC - Matthews correlation coefficient; HSS - Heidke skill score.
}
\end{deluxetable*}

\section{Conclusions} \label{sec:conclusion}
Signatures of solar activity constitute many transient events, including solar flares, coronal mass ejections, and solar energetic particles. These are the main drivers of space weather. Towards predictive efforts of SEP events, we have built and analyzed the GSEP data set first presented in \citet{rotti2022} and further discussed in this paper. Our extended data set consists of 433 events out of which 244 cross the 10 pfu threshold in the E$\geq$10MeV proton channel. We rely on several sources such as existing event catalogs of solar flares, CMEs, radio bursts and SEP events, and relevant observational data from SOHO and SDO missions to associate each SEP event in our data set with a flare and/or CME wherever possible.

We conduct statistical analysis of SEP events along with the parent solar eruptions, namely, estimating the relationships of flare magnitude, its rise time, flaring locations, speed and width of CMEs, and temporal variations of sources with SEP events. With respect to SEP peak fluxes and event fluences, we find a good correlation for CME parameters, and moderate correlations are observed for X-ray flare intensities. Most of the large SEP events in our data set are associated with intense flares ($\geq$M2.0). Also, we find most (66\%) of the CMEs associated with SEP events are halo in angular extents and fast with a median speed of 1200 km.s$^{-1}$. In summary, as shown in previous works such as \cite{cane2010study} and \cite{papaioannou2016solar}, the SEP event intensities increase with increasing CME speed and flare strength. The distribution of event origins shows an increase in proportionality with source eruptions occurring on the visible disk up to western hemisphere. Exceptions are noticed for very large flare sources ($\geq$X1.0) occurring over the eastern and beyond-limb locations due to poor magnetic connectivity for the particles to reach Earth.

We focus on extracting important features in our data set by implementing tree-based machine learning models, namely, random forest and XGBoost classifiers. We find both the CME speed and flare intensity/magnitude to be the top contributing features towards the corresponding SEP event peak fluxes. In a recent study, \cite{lavasa2021assessing} undertake a feature extraction process and find CME speed, width and flare soft x-ray fluence as the most important features for identification of SEP events.

In this work, we also implement logistic regression and support vector machines in the framework of a binary classification problem. We consider large SEP events as the positive class and weak ones as the negative. We perform hyperparameter tuning and implement a 2-fold cross-validation technique to optimize our models for robustness. We use F1 score, Matthews correlation coefficient (MCC), true skill statistics (TSS) and Heidke skill score (HSS) to assess the model performance. The results in this paper show that all our models perform moderately well. For tree-based models, we see XGBoost perform slightly better than RF while LR and SVM are close in their outcomes. There are many areas in which SEP event prediction efforts can be improved and we emphasize undertaking further study on feature engineering in future work.

The GSEP data set is available at Harvard Dataverse repository at \dataset[10.7910/DVN/DZYLHK]{\doi{10.7910/DVN/DZYLHK}}. The statistical plots and analysis presented in this study are based on version 5.0 of the data set.

\begin{acknowledgments}
We acknowledge the use of X-Ray and proton flux data from the GOES missions made available by NOAA. We also acknowledge the use of observations from SOHO and SDO missions and thank the team for the availability of the data online. SOHO is a project of international cooperation between ESA and NASA. SDO is the first mission to be launched for NASA's Living With a Star (LWS) Program. We also thank the CDAW for the opportunity to utilize their SEP, CME and type-II radio burst catalogs. We thank the anonymous reviewer for constructive comments on the manuscript that have improved the contents of the paper. Author Petrus Martens' contribution is supported by NASA SWR2O2R Grant 80NSSC22K0272. Author S. Rotti carried out this work while supported by the NASA FINESST Grant 80NSSC21K1388.

\end{acknowledgments}
%



\software{pandas \citep{mckinney2010data}, numpy \citep{van2011numpy, harris2020array}, sklearn \citep{scikit-learn}, xgboost \citep{XGB}, matplotlib \citep{hunter2007matplotlib}, seaborn \citep{waskom_2017_883859, waskom2021seaborn}}



\appendix
\section{GOES Data}
During the data collection and preparation stage, we also obtained a list of GOES primary and secondary missions from SWPC. The reported measurements of peak fluxes are not consistent over a few SEP events across the existing catalogs. Hence, we visually inspected all the GOES proton data from 1986 to 2017 to understand the differences and also the variations in measurements between parallel GOES missions. We compiled a list of the GOES missions from -5 to -15 during the last three solar cycles from which we considered the proton flux data for the present work. The list is presented here in table \ref{tab:goes} below. The underlined text indicates a primary satellite from the GOES mission as identified by NOAA.
 
\begin{table}[ht!]
\tablenum{4}
\caption{List of GOES missions considered in the present work.}
\label{tab:goes}
\begin{tabular}{lllllllllllll}
     & Jan       & Feb       & Mar       & Apr       & May       & Jun       & Jul       & Aug       & Sep       & Oct       & Nov       & Dec       \\
\hline
1986 & {\ul G06} & G05       & G05       & {\ul G06} & G05       & {\ul G06} & {\ul G06} & {\ul G06} & {\ul G06} & {\ul G06} & {\ul G06} & {\ul G06} \\
1987 & {\ul G06} & G05       & {\ul G06} & {\ul G06} & {\ul G06} & {\ul G06} & {\ul G06} & {\ul G06} & G06       & {\ul G06} & {\ul G06} & {\ul G06} \\
1988 & {\ul G06} & {\ul G07} & {\ul G07} & {\ul G07} & {\ul G07} & {\ul G07} & {\ul G07} & G06       & {\ul G07} & {\ul G07} & {\ul G07} & {\ul G07} \\
1989 & {\ul G07} & {\ul G07} & {\ul G07} & {\ul G07} & G06       & {\ul G07} & {\ul G07} & {\ul G07} & {\ul G07} & {\ul G07} & {\ul G07} & {\ul G07} \\
1990 & {\ul G07} & {\ul G07} & {\ul G07} & {\ul G07} & {\ul G07} & G06       & {\ul G07} & {\ul G07} & {\ul G07} & {\ul G07} & {\ul G07} & {\ul G07} \\
1991 & {\ul G07} & {\ul G07} & {\ul G07} & {\ul G07} & {\ul G07} & {\ul G07} & {\ul G07} & {\ul G07} & {\ul G07} & {\ul G07} & {\ul G07} & {\ul G07} \\
1992 & {\ul G07} & {\ul G07} & {\ul G07} & {\ul G07} & G06       & {\ul G07} & {\ul G07} & {\ul G07} & {\ul G07} & {\ul G07} & {\ul G07} & {\ul G07} \\
1993 & {\ul G07} & {\ul G07} & {\ul G07} & {\ul G07} & {\ul G07} & {\ul G07} & {\ul G07} & {\ul G07} & {\ul G07} & {\ul G07} & {\ul G07} & {\ul G07} \\
1994 & {\ul G07} & {\ul G07} & {\ul G07} & {\ul G07} & {\ul G07} & {\ul G07} & {\ul G07} & {\ul G07} & {\ul G07} & {\ul G07} & {\ul G07} & {\ul G07} \\
1995 & {\ul G08} & {\ul G08} & {\ul G08} & {\ul G08} & {\ul G08} & {\ul G08} & {\ul G08} & {\ul G08} & {\ul G08} & {\ul G08} & {\ul G08} & {\ul G08} \\
1996 & {\ul G08} & {\ul G08} & {\ul G08} & {\ul G08} & {\ul G08} & {\ul G08} & {\ul G08} & {\ul G08} & {\ul G08} & {\ul G08} & {\ul G08} & {\ul G08} \\
1997 & {\ul G08} & {\ul G08} & {\ul G08} & {\ul G08} & {\ul G08} & {\ul G08} & {\ul G08} & {\ul G08} & {\ul G08} & {\ul G08} & {\ul G08} & {\ul G08} \\
1998 & {\ul G08} & {\ul G08} & {\ul G08} & {\ul G08} & {\ul G08} & {\ul G08} & {\ul G08} & G10       & G10       & {\ul G08} & {\ul G08} & G10       \\
1999 & {\ul G08} & {\ul G08} & {\ul G08} & {\ul G08} & {\ul G08} & {\ul G08} & {\ul G08} & {\ul G08} & {\ul G08} & {\ul G08} & {\ul G08} & {\ul G08} \\
2000 & {\ul G08} & {\ul G08} & {\ul G08} & {\ul G08} & {\ul G08} & {\ul G08} & {\ul G08} & {\ul G08} & {\ul G08} & {\ul G08} & {\ul G08} & {\ul G08} \\
2001 & {\ul G08} & {\ul G08} & {\ul G08} & {\ul G08} & {\ul G08} & {\ul G08} & G10       & G10       & G10       & G10       & G10       & G10       \\
2002 & G10       & G10       & G10       & {\ul G08} & {\ul G08} & {\ul G08} & G10       & G10       & G10       & G10       & G10       & G10       \\
2003 & G10       & G10       & G10       & {\ul G10} & {\ul G10} & G10       & {\ul G11} & {\ul G11} & {\ul G11} & {\ul G11} & {\ul G11} & {\ul G11} \\
2004 & {\ul G11} & {\ul G11} & {\ul G11} & {\ul G11} & {\ul G11} & {\ul G11} & {\ul G11} & {\ul G11} & {\ul G11} & {\ul G11} & {\ul G11} & {\ul G11} \\
2005 & {\ul G11} & {\ul G11} & {\ul G11} & {\ul G11} & {\ul G11} & {\ul G11} & {\ul G11} & {\ul G11} & {\ul G11} & {\ul G11} & {\ul G11} & {\ul G11} \\
2006 & {\ul G11} & {\ul G11} & {\ul G11} & {\ul G11} & {\ul G11} & {\ul G11} & {\ul G11} & {\ul G11} & {\ul G11} & {\ul G11} & {\ul G11} & {\ul G11} \\
2007 & {\ul G11} & {\ul G11} & {\ul G11} & {\ul G11} & {\ul G11} & {\ul G11} & {\ul G11} & {\ul G11} & {\ul G11} & {\ul G11} & {\ul G11} & {\ul G11} \\
2008 & {\ul G11} & {\ul G11} & {\ul G11} & {\ul G11} & {\ul G11} & {\ul G11} & {\ul G11} & {\ul G11} & {\ul G11} & {\ul G11} & {\ul G11} & {\ul G11} \\
2009 & {\ul G11} & {\ul G11} & {\ul G11} & {\ul G11} & {\ul G11} & {\ul G11} & {\ul G11} & {\ul G11} & {\ul G11} & {\ul G11} & {\ul G11} & {\ul G11} \\
2010 & {\ul G11} & {\ul G11} & {\ul G11} & {\ul G11} & G11       & G11       & G11       & G11       & G11       & G11       & G11       & G11       \\
2011 & G11       & G11       & G15       & G15       & G15       & G15       & G15       & G15       & {\ul G13} & {\ul G13} & G15       & G15       \\
2012 & G15       & G15       & G15       & {\ul G13} & G15       & G15       & G15       & G15       & G15       & {\ul G13} & {\ul G15} & {\ul G15} \\
2013 & {\ul G15} & {\ul G15} & {\ul G15} & {\ul G13} & {\ul G15} & G15       & G15       & G15       & {\ul G13} & G15       & {\ul G13} & {\ul G13} \\
2014 & {\ul G13} & {\ul G13} & {\ul G13} & G15       & G15       & G15       & G15       & G15       & G15       & G15       & G15       & G15       \\
2015 & {\ul G13} & {\ul G13} & G15       & G15       & G15       & {\ul G13} & {\ul G13} & {\ul G13} & {\ul G13} & G15       & {\ul G13} & {\ul G13} \\
2016 & G15       & G15       & G15       & G15       & G15       & G15       & G15       & G15       & G15       & G15       & G15       & G15       \\
2017 & G15       & G15       & G15       & G15       & G15       & {\ul G15} & {\ul G13} & G15       & G15       & G15       & G15       & G15
\end{tabular}

\tablecomments{`G' stands for GOES and the corresponding number indicates the mission. For example, `G15' represents the GOES-15. Underlined names of satellites indicate the `primary' designation by NOAA.}
\end{table}


\bibliography{sample631}{}
\bibliographystyle{aasjournal}


\end{document}